\documentclass[aos,preprint]{imsart}
\setattribute{journal}{name}{}
\RequirePackage[round]{natbib}
\RequirePackage[colorlinks,citecolor=blue,urlcolor=blue]{hyperref}

\usepackage{pdfpages}

\usepackage{bbm}
\usepackage{comment}
\usepackage{amsmath, amsthm, amssymb}
\usepackage{algpseudocode}
\usepackage{algorithm}
\usepackage{graphicx}
\usepackage{subfigure}
\usepackage{lineno}
\usepackage[left=3cm,right=3cm, bottom=3cm]{geometry}

\let\hat\widehat

\theoremstyle{remark}

\newcommand\R{\mathbb{R}}
\newcommand\K{\mathbb{K}}
\newcommand\E{\mathbb{E}}

\newcommand\cR{{\cal R}}

\newcommand\cE{{\cal E}}

\catcode`@=11
\newskip\beforeproofvskip
\newskip\afterproofvskip
\beforeproofvskip=\medskipamount
\afterproofvskip=\bigskipamount

\def\prooftag{Proof}
\def\proofskip{\enspace}

\def\proof{\@ifnextchar[{\@@proof}{\@proof}}  
\def\@startproof{\par\vskip\beforeproofvskip\leavevmode}
\def\@proof{\@startproof{\scshape\prooftag.}\proofskip}
\def\@@proof[#1]{\@startproof {\scshape\prooftag #1.}\proofskip}

\catcode`@=12

\let\hat\widehat
\let\tilde\widetilde

\begin{document}

%
%
%
%
%
%
%
%
%
%

\begin{center}
{\bf\Large A Tutorial on Kernel Density Estimation and Recent Advances}\\ \vspace{.2cm}
{\bf 
Yen-Chi Chen\\
Department of Statistics\\
University of Washington\\
{\bf\today}
}
\end{center}

\begin{quote}
{\em This tutorial provides a gentle
introduction to kernel density estimation (KDE) and recent advances regarding confidence bands
and geometric/topological features. 
We begin with a discussion of basic properties of KDE:
the convergence rate under various metrics,
density derivative estimation, and bandwidth selection.
Then, we introduce common approaches to the construction of
confidence intervals/bands, and 
we discuss how to handle bias. 
Next, we talk about recent advances in the inference of geometric and topological features of a density function
using KDE. 
Finally, we illustrate how one can use KDE
to estimate a cumulative distribution function and a receiver operating characteristic curve. 
We provide R implementations related to this tutorial at the end.
}
\end{quote}

\begin{quote}
\end{quote}


\tableofcontents

\pagebreak

\section{Introduction}		\label{sec::intro}

Kernel density estimation (KDE), also known as the Parzen's window \citep{parzen1962estimation}, 
is one of the most well-known approaches to estimate the underlying probability density function
of a dataset. 
KDE is a nonparametric density estimator requiring no assumption that
the underlying density function is from a parametric family.
KDE will learn the shape of the density from the data automatically.
This flexibility arising from its nonparametric nature makes KDE a very popular approach 
for data drawn from a complicated distribution.

\begin{figure}
\center
\includegraphics[width=2.25in]{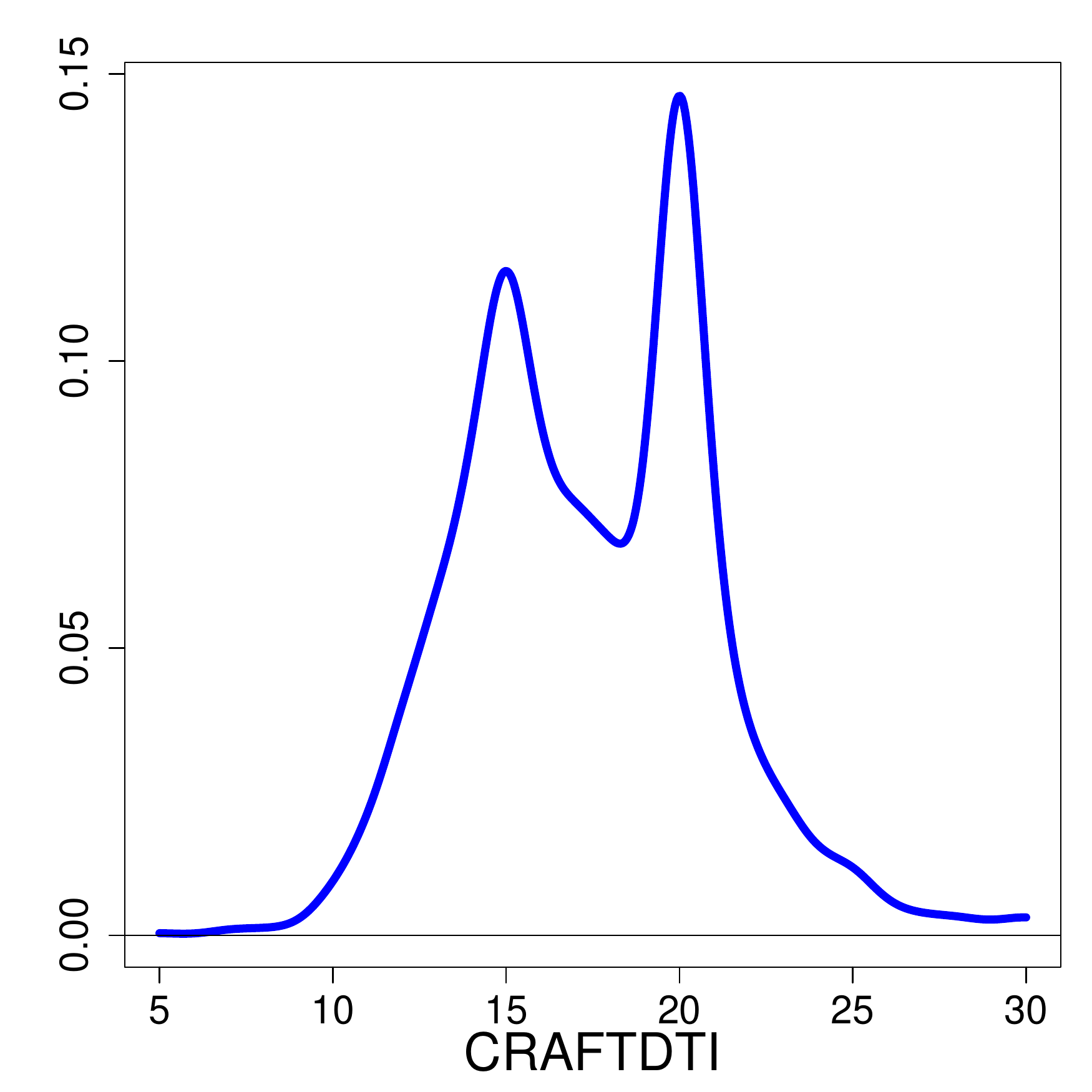}
\includegraphics[width=2.25in]{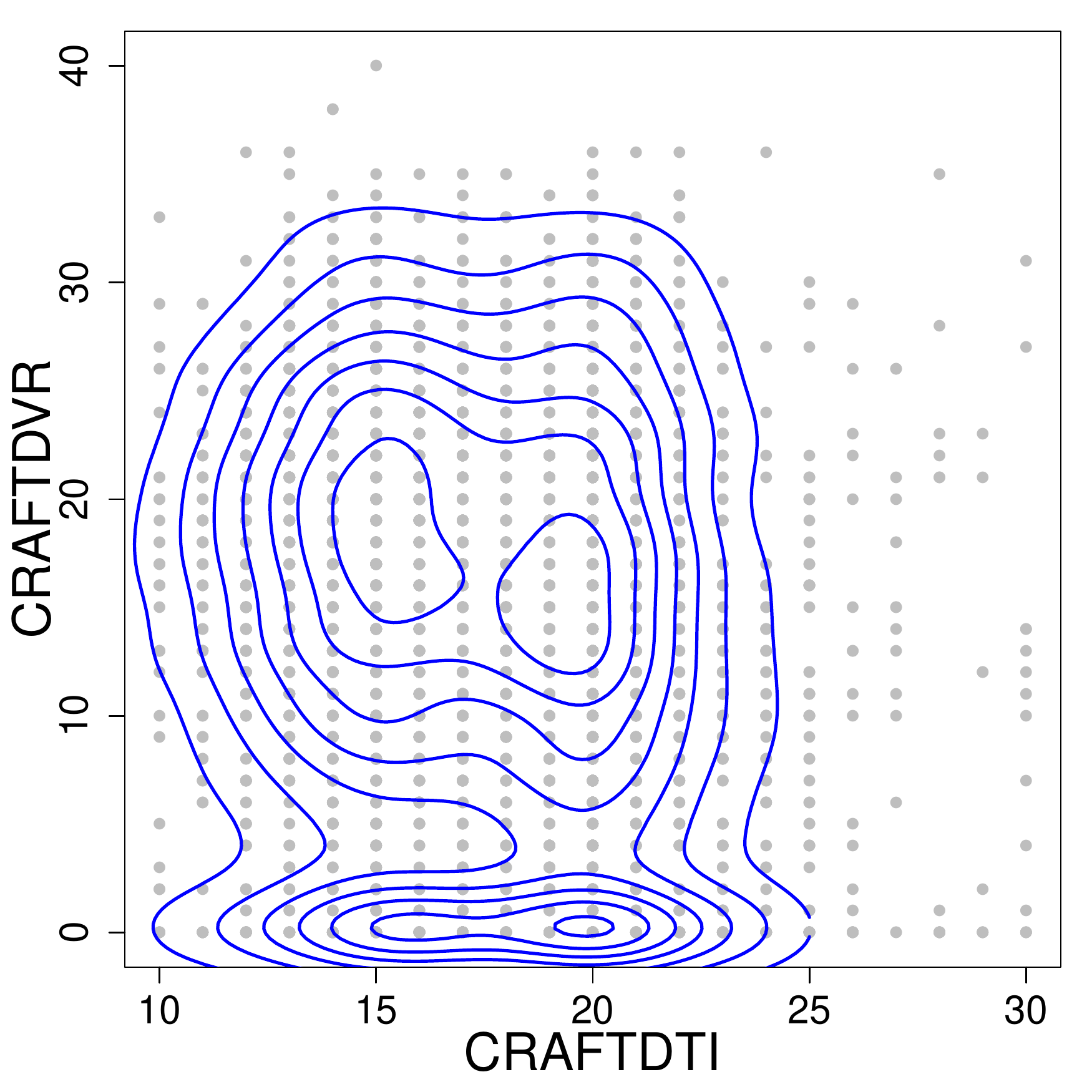}
\caption{Examples of KDE using the NACC Uniform Data Set.
We focus on variables `CRAFTDTI' and `CRAFTDVR' and
use those subjects who have non-missing value of either variables.
{\bf Left:} 
We show the marginal density function of variable `CRAFTDTI'
from one-dimensional (1D) KDE.
There are two bumps in for this density function. 
{\bf Right:} 
We show the scatter plot along with the bivariate density function of both variables from the 
two-dimensional (2D) KDE,
showing the multimodality feature of this bivariate density function.
}
\label{fig::ex0}
\end{figure}

Figure~\ref{fig::ex0} illustrates KDE using a part of the NACC (National AlzheimerÕs Coordinating Center) 
Uniform Data Set \citep{beekly2007national}, version 3.0 (March 2015).
Because the purpose of using this dataset is to illustrate the effectiveness of KDE, 
we will draw no scientific conclusion
but will just use KDE as a tool to explore the pattern of the data.
We focus on two variables, `CRAFTDTI' (Craft Story 21 Recall -- delay time),
and `CRAFTDVR' (Craft Story 21 Recall -- total story units recalled, verbatim scoring). 
Although these two variables take integer values, we treat them as continuous
and use KDE to determine the density function.
We consider only the unique subject with scores on both variables, resulting in
a sample of size $4,044$.
In the left panel of Figure~\ref{fig::ex0}, we display the estimated density function of `CRAFTDTI'
using KDE. 
We see that there are two modes in the distribution. 
In the right panel of Figure~\ref{fig::ex0}, we show the scatter plot of the data and overlay it with the 
result of bivariate KDE 
(blue contours). 
Because many subjects have identical values for the two variables, the scatter plot (gray dots) 
provides no useful information
regarding the underlying distribution. 
However, KDE shows the multi-modality of this bivariate distribution,
which contains multiple bumps that cannot be captured easily by any parametric distribution.


%

The remainder of the tutorial is organized as follows.
In Section~\ref{sec::KDE},
we present the definition of KDE, 
followed by a discussion of its basic properties:
convergence rates, density derivative estimations, and bandwidth selection. 
Then, in Section~\ref{sec::CS},
we introduce common approaches to the construction of confidence regions,
and we discuss the problem of bias in statistical inference.
Section~\ref{sec::GT} provides an introduction to the use of
KDE to estimate geometric and topological features of a 
density function.
In Section~\ref{sec::CDF}, we study how one can use KDE
to estimate the cumulative distribution function (CDF) and the receiver operating characteristic (ROC) curve. 
Finally, in Section~\ref{sec::future}, we discuss open problems.
At the end of this tutorial, we provide R codes
for implementing the presented analysis of KDE.

\section{Statistical Properties}	\label{sec::KDE}

Let $X_1,\cdots, X_n\in\R^d$ be an independent, identically distributed random sample from an unknown 
distribution $P$ with density function $p$.
Formally, KDE can be expressed as
\begin{equation}
\hat{p}_n(x) = \frac{1}{nh^d}\sum_{i=1}^n K\left(\frac{x-X_{i}}{h}\right),
\label{eq::KDE}
\end{equation}
where $K: \R^d \mapsto \R$ is a smooth function called the kernel function and $h>0$ is the smoothing bandwidth 
that controls the amount of smoothing. 
Two common examples of $K(x)$ are 
\begin{align*}
\mbox{(Gaussian kernel)}\quad K(x) &= \frac{\exp\left(-\|x\|^2/2\right)}{v_{1,d}},\quad v_{1,d}= \int \exp\left(-\|x\|^2/2\right)dx,\\
\mbox{(Spherical kernel)}\quad K(x) &= \frac{I(\|x\|\leq 1)}{v_{2,d}},\quad v_{2,d}= \int I(\|x\|\leq1) dx.
\end{align*}
Note that we apply the same amount of smoothing $h$ in every direction;
in practice, one can use a bandwidth matrix $H$ 
and the quantity $K\left(\frac{x-X_{i}}{h}\right)$ becomes $K\left(H^{-1} (x-X_i)\right)$.



Intuitively, KDE has the effect of smoothing out each data point into a smooth bump, 
whose the shape
is determined by the kernel function $K(x)$.
Then, KDE sums over all these bumps to obtain a density estimator.
At regions with many observations, because there will be many bumps around,
KDE will yield a large value.
On the other hand, for regions with only a few observations, the density value from summing over the bumps
is low, because only have a few bumps contribute to the density estimate.


\begin{figure}
\center
\includegraphics[width=2.25in]{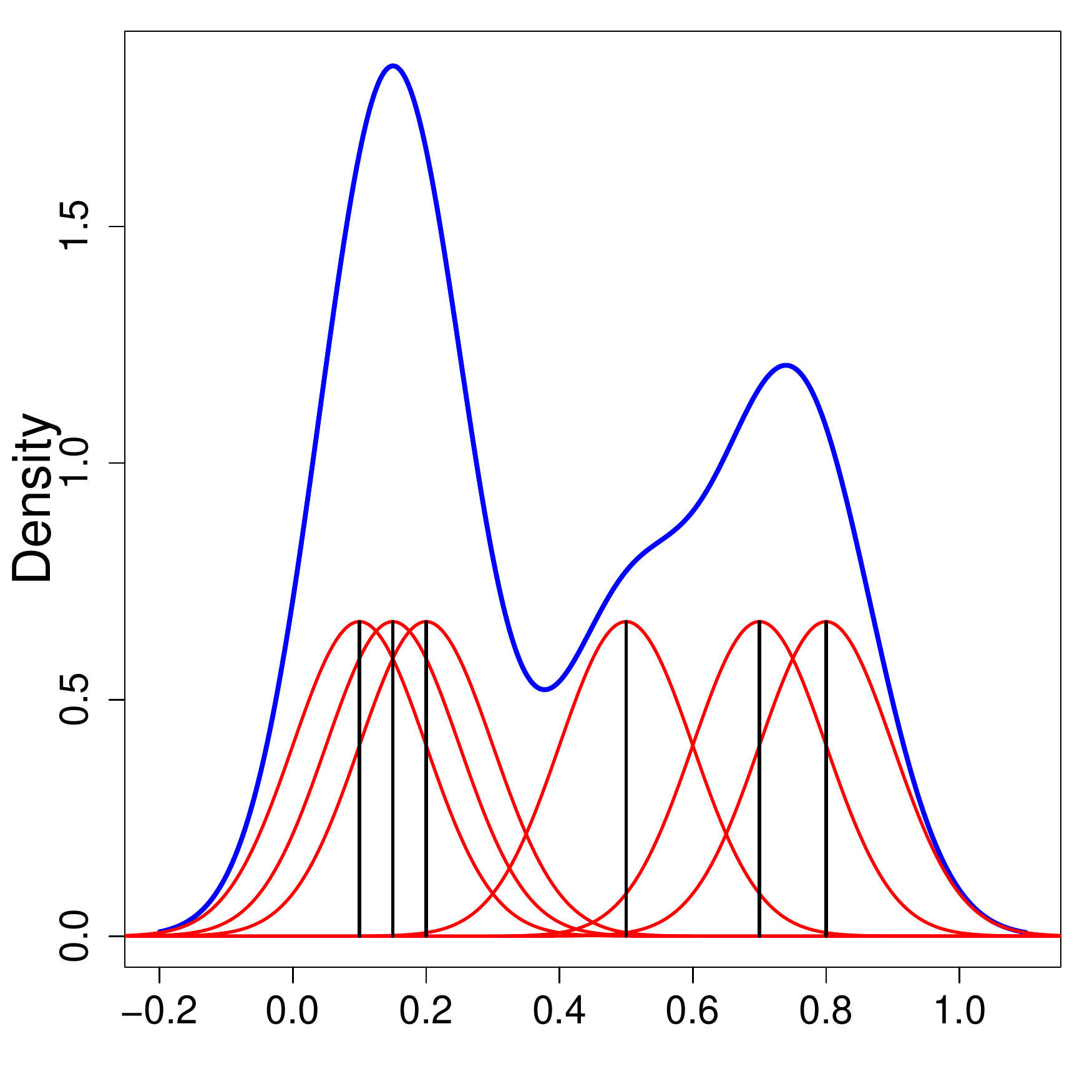}
\caption{
A 1D illustration of how KDE is constructed. 
There are six observations, located at the positions indicated by black lines. 
We then smooth these observations into small bumps (red bumps)
and the sum them to obtain the density estimate (blue curve). 
}
\label{fig::ex}
\end{figure}

Figure~\ref{fig::ex} presents examples of KDE in the 1D case.
There are six observations, as indicated by the black lines.
We smooth these observations into bumps (red bumps) and
sum over all of the bumps to form the final density estimator (the blue curve).
In R, many packages are equipped with programs for computing KDE;
see
\cite{deng2011density} for a listing.


%
%

{\bf Remark.}
(Adaptive smoothing)
The amount of smoothing can depend on the location $x$ \citep{loftsgaarden1965nonparametric} or
the data point $X_i$ \citep{breiman1977variable}.
In the former case, 
we use 
$h=h(x)$ so KDE becomes
$
\hat{p}_n(x) = \frac{1}{nh^d(x)}\sum_{i=1}^n K\left(\frac{x-X_{i}}{h(x)}\right),
$
which is referred to as the balloon estimator \citep{terrell1992variable}. 
In the latter case, 
we use $h = h_i =h(X_i)$ for the $i$-th data points with the resulting density estimate being
$
\hat{p}_n(x) = \frac{1}{n}\sum_{i=1}^n \frac{1}{h_i^d}K\left(\frac{x-X_{i}}{h_i}\right),
$
which is referred to as the sample smoothing estimator \citep{terrell1992variable}.
For more details regarding adaptive smoothing, we refer the readers to Section 6.6 of \cite{scott2015multivariate}.

\subsection{Convergence Rate}	\label{sec::CR}
To measure the errors of KDE, we consider three types of errors:
the pointwise error, uniform error, and mean integrated square error (MISE).
The pointwise error is the simplest error and is related to the confidence interval (Section~\ref{sec::PCS}).
The uniform error has many useful theoretical properties since it measures the uniform deviation of the estimator
and can be used to bound other types of errors.
The uniform error is related to the confidence band and 
the geometric (and topological) features; see
Section~\ref{sec::UCS} and Section~\ref{sec::GT}.
The MISE (actually it is a risk measurement of the estimator)
is generally used in bandwidth selection (Section~\ref{sec::BS}), because it measures the overall
performance of the estimator and is related to the mean square error.

{\bf Pointwise Error.}
For a given point $x$, the pointwise error of KDE
is the difference between KDE $\hat{p}_n(x)$ and $p(x)$, the true density function evaluated at $x$.
Let $\nabla^2 p = \sum_{\ell=1}^d\frac{\partial^2p}{\partial x^2_\ell}$ be the Laplacian of the function $p$.
Under smoothness conditions \citep{scott2015multivariate,wasserman2006all,tsybakov2009introduction}, 
\begin{equation}
\begin{aligned}
\hat{p}_n(x)- p(x)& = \underbrace{\E\left( \hat{p}_n(x)\right)- p(x)}_{B_h(x)} +\underbrace{\hat{p}_n(x)-\E\left( \hat{p}_n(x)\right)}_{\cE_n(x)}\\
& = O(h^2) + O_P\left(\sqrt{\frac{1}{nh^d}}\right),\\
B_h(x) & = \frac{h^2}{2} \sigma_K^2 \nabla^2 p(x) + o(h^2),\\
\cE_n(x) & = \sqrt{\frac{\mu_K \cdot p(x)}{nh^d}}\cdot Z_n(x) +o_P\left(\sqrt{\frac{1}{nh^d}}\right),
\end{aligned}
\label{eq::PE}
\end{equation}
where $Z_n(x)\overset{D}{\rightarrow} N(0, 1)$ and $\sigma_K^2= \int \|x\|^2 K(x)dx, \mu_K = \int K^2(x)dx$ 
are constants depending only on the kernel function $K$.
Thus, when $h\rightarrow 0$ and $nh^d \rightarrow \infty$, $\hat{p}_n(x)\overset{P}{\rightarrow} p(x)$, i.e.,
KDE $\hat{p}_n(x)$ is a consistent estimator of $p(x)$.
Equation \eqref{eq::PE} presents the decomposition of
the (pointwise) estimation error of KDE in terms of the bias $B_h(x)$ 
and the stochastic variation $\cE_n(x)$.
This decomposition will be used frequently in deriving other errors
and constructing the confidence regions.





{\bf Uniform Error.}
Another error metric is the uniform error (also known as the $L_\infty$ error),
the maximal difference between $\hat{p}_n$ and $p$: $\sup_x|\hat{p}_n(x)- p(x)|$.
According to empirical process theory 
\citep{yukich1985laws,Gine2002,Einmahl2005,rao2014nonparametric}, 
the uniform error
\begin{equation}
\begin{aligned}
\sup_x|\hat{p}_n(x)- p(x)|
& = O(h^2) + O_P\left(\sqrt{\frac{\log n}{nh^d}}\right)
\end{aligned}
\label{eq::UE}
\end{equation}
under mild conditions (see \cite{Gine2002,Einmahl2005} for more details).
The error rate in \eqref{eq::UE} and the pointwise error rate in \eqref{eq::PE}
differ only in the stochastic variation portion
and the difference is at the rate $\sqrt{\log n}$.
The presence of an extra $\sqrt{\log n}$ in the uniform error rate is a very common phenomenon 
in nonparametric estimation
owing to empirical process theory.
The uniform error has many useful theoretical properties
\citep{chen2015density,chen2016generalized,fasy2014confidence,jisu2016statistical},
because it provides a uniform control of the estimation error over the entire support.

{\bf MISE.}
The MISE is one of the most
well-known error measurements \citep{wasserman2006all, scott2015multivariate}
among all of the error measures used in KDE.
The MISE is defined as $\int \E\left(\left(\hat{p}_n(x)-p(x)\right)^2\right) dx$.
Thus, the MISE measures the $L_2$ risk of KDE.
Under regularity conditions, the MISE
\begin{equation}
\begin{aligned}
\int \E\left(\left(\hat{p}_n(x)-p(x)\right)^2\right) dx & = 
\int B^2_h(x)dx + \int {\sf Var}(\cE_n(x)) dx\\
&=
\frac{h^4}{4}\sigma_K^4 \int\left| \nabla^2 p(x)\right|^2dx + \frac{\mu_K}{nh^d} + o(h^4) + o\left(\frac{1}{nh^d}\right).
\end{aligned}
\label{eq::MISE}
\end{equation}
The MISE can be viewed as the mean square error of KDE. 
The dominating term $\frac{h^4}{4}\sigma_K^4 \int\left| \nabla^2 p(x)\right|^2dx+ \frac{\mu_K}{nh^d}$
is called the asymptotic mean integrated square error (AMISE).
Equation \eqref{eq::MISE} shows that the error (risk) of KDE
can be decomposed in to a bias component, $\frac{h^4}{4}\sigma_K^4 \int\left| \nabla^2 p(x)\right|^2dx$, 
and a variance component $\frac{\mu_K}{nh^d}$
together with small corrections.
This decomposition is known as the bias-variance tradeoff \citep{wasserman2006all} and 
is very useful in practice because we can choose
the smoothing bandwidth $h$ by optimizing this error.
If we ignore smaller order terms and use the AMISE,
the minimal error occurs when we choose
\begin{equation}
h_{\sf opt}  = \left(\frac{4 \mu_K}{\sigma_K^4 \int\left| \nabla^2 p(x)\right|^2dx}\cdot \frac{1}{n}\right)^{\frac{1}{d+4}} 
\label{eq::h}
\end{equation}
which leads to the optimal MISE
$$
\int \E\left(\left(\hat{p}_{n, {\sf opt}}(x)-p(x)\right)^2\right) dx = \inf_{h>0}\int \E\left(\left(\hat{p}_{n}(x)-p(x)\right)^2\right) dx  = O\left(n^{-\frac{2}{d+4}}\right).
$$
Equation \eqref{eq::h} will be a key result in choosing the smoothing bandwidth (Section \ref{sec::BS}). 
Note that in practice, people generally select the smoothing bandwidth by minimizing the 
MISE rather than other errors because
(i) it is a risk function that does not depend on any particular sample, 
(ii) it measures the overall estimation error rather than putting
too much weight on a small portion of the support (i.e., it is more robust to small perturbations), and
(iii) it has useful theoretical behaviors, including the expression of the bias-variance tradeoff
and the connection to the mean square error.

{\bf Remark.} (Boundary bias)
When the density function is discontinuous, the bias of KDE at the discontinuities 
will be of the order $O(h)$ rather than $O(h^2)$, and this bias is called the 
boundary bias \citep{wasserman2006all, scott2015multivariate}.
In practice, one can use the boundary kernel to reduce the boundary bias 
(see, e.g., Chapter 6.2.3.5 in \cite{scott2015multivariate}).

\subsection{Derivative Estimation}	\label{sec::deriv}

KDE can be used to estimate the derivative of the density function.
This is often called \emph{density derivative estimation} \citep{stoker1993smoothing,chacon2011asymptotics}.
The idea is simple: we use
the derivative of KDE as an estimator of the corresponding derivative of the density function.
Let $[\beta] = (\beta_1,\cdots,\beta_d)$ be a multi-index (each $\beta_\ell$ is a non-negative integer
and $|[\beta]|= \sum_{\ell=1}^d \beta_d$).
Define $D^{[\beta]} = \frac{\partial^{\beta_1}}{\partial x_1^{\beta_1}}\cdots\frac{\partial^{\beta_d}}{\partial x_d^{\beta_d}}$
to be the $[\beta]$-th order partial derivative operator.
For instance, $[\beta] = (1,3,0,\cdots,0)$ implies $D^{[\beta]} =\frac{\partial}{\partial x_1}\frac{\partial^{3}}{\partial x_2^{3}}$
and $|[\beta]|=  4$.
Then, under smoothness assumptions \citep{chacon2011asymptotics}, 
\begin{equation}
D^{[\beta]}\hat{p}_n(x) - D^{[\beta]}p(x) = O(h^2) + O_P\left(\sqrt{\frac{1}{nh^{d+2|[\beta]|}}}\right).
\label{eq::Drate}
\end{equation}
That is, the (MISE or pointwise) error rate of gradients of KDE is $O(h^2) + O_P\left(\sqrt{\frac{1}{nh^{d+2}}}\right)$
and the error rate of second derivatives (Hessian matrix) of KDE is $O(h^2) + O_P\left(\sqrt{\frac{1}{nh^{d+4}}}\right)$.
Similarly to $B_h(x)$ and $\cE_n(x)$ in the density estimation,
there are explicit formulas for
the bias and stochastic variation of density derivative estimation;
see \cite{chacon2011asymptotics} for more details. 
Some examples of using gradient and second derivative estimation can be found in 
\cite{arias2016estimation, chacon2013data,chen2015asymptotic,chen2015density,genovese2009path,genovese2014nonparametric}.

\begin{figure}
\center
\includegraphics[width=1.8in]{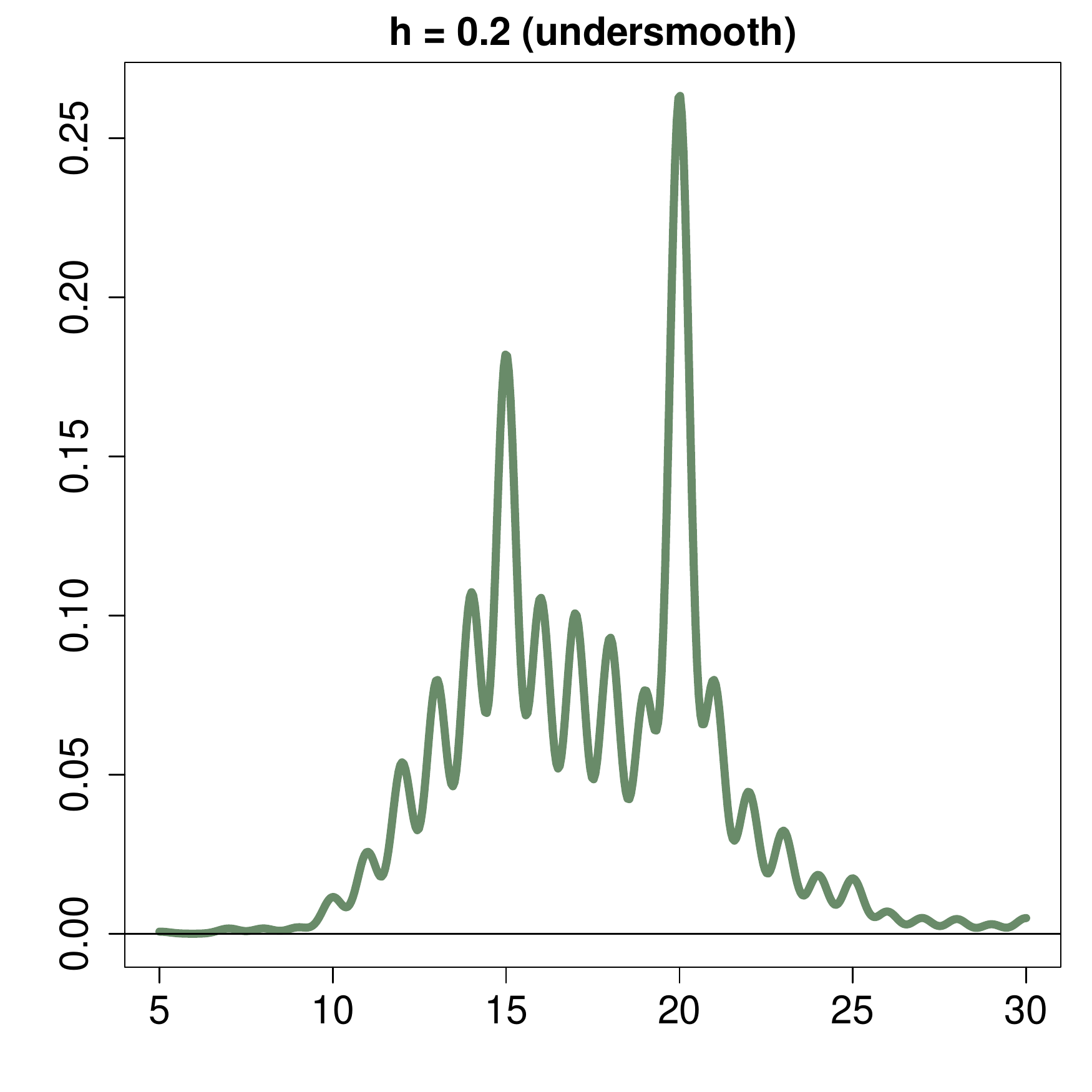}
\includegraphics[width=1.8in]{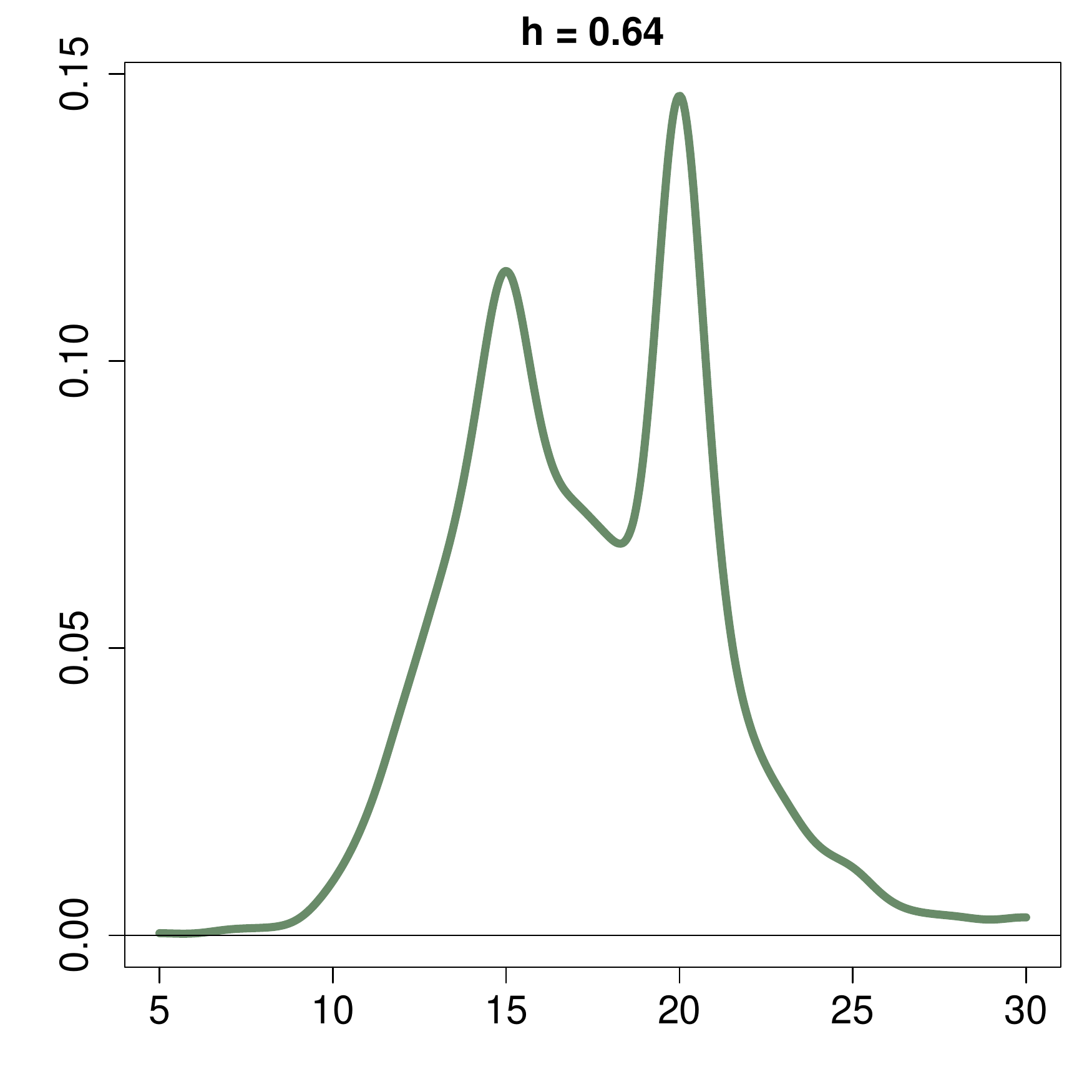}
\includegraphics[width=1.8in]{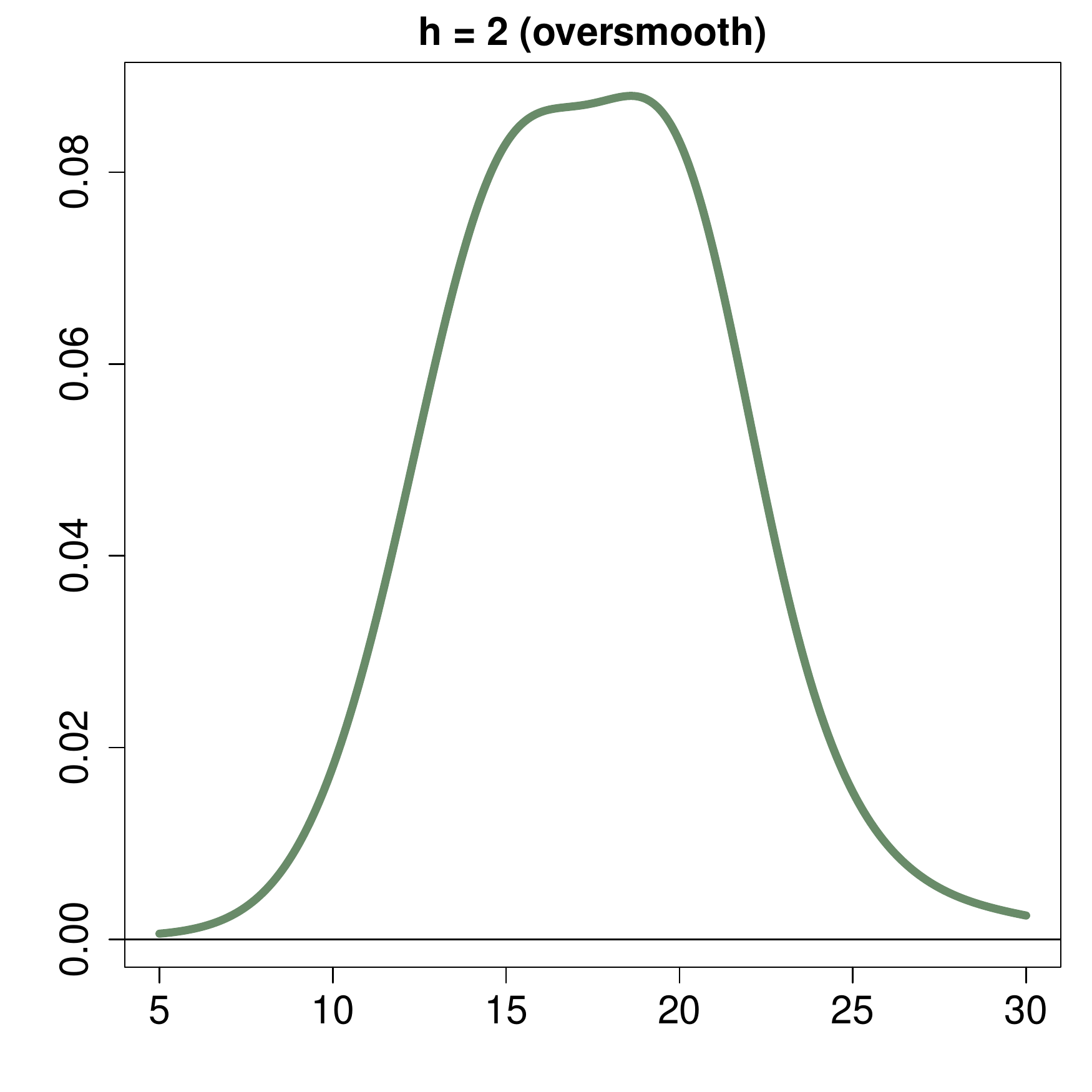}
\caption{Smoothing bandwidth and KDE.
We use the same data as in the left panel of
Figure~\ref{fig::ex0}.
We display KDE using three different amounts of smoothing.
The left panel is the case of undersmoothing: we choose an excessively small bandwidth $h$.
The middle panel is the case of the correct amount of smoothing, which is chosen 
according to the default rule in R.
The right panel is the case of oversmoothing: the chosen $h$ is too large.
}
\label{fig::BW}
\end{figure}

\subsection{Bandwidth Selection}	\label{sec::BS}

How to choose the smoothing bandwidth for KDE
is a classical research topic in nonparametric statistics.
This problem is often known as \emph{bandwidth selection}. 
Figure~\ref{fig::BW} shows KDE's with 
different amounts of smoothing of the same dataset.
When $h$ is too small (left panel), there are many wiggles in the density estimate.
When $h$ is too large (right panel), we smooth out important features. 
When $h$ is at the correct amount (middle), we can see a clear picture of the underlying density.

Common approaches to bandwidth selection include
the
rule of thumb \citep{Silverman1986},
least square cross-validation,
\citep{rudemo1982empirical, bowman1984alternative, bowman1997applied,stone1984asymptotically},
biased cross-validation,
\citep{scott1987biased},
and plug-in method \citep{woodroofe1970choosing, sheather1991reliable}.
Roughly speaking, the core idea behind all of these methods is to minimize the 
AMISE, the dominating 
quantity in the MISE \eqref{eq::MISE},
or other similar error measurements.
Different bandwidth selectors can be viewed as different estimators to the AMISE,
and $h$ is chosen by minimizing the AMISE estimator.
Overviews and comparisons of the existing methods
can be found in 
\cite{jones1996brief, sheather2004density},
page 135--137 in \cite{wasserman2006all},
and Chapter 6.5 in
\cite{scott2015multivariate}.

While most of the literature focuses on the univariate case,
\cite{chacon2013data} provides a generalization of all of the above methods
to the multivariate case and also generalizes the AMISE criterion into density derivative estimation. 
In R, one can use packge `{\sf ks}\footnote{\url{https://cran.r-project.org/web/packages/ks/index.html}}' or 
`{\sf kedd}\footnote{\url{https://cran.r-project.org/web/packages/kedd/index.html}}'
to choose smoothing bandwidths
for both density estimation and density derivative estimation.
Note that the {\sf ks} package is applicable to multivariate data.

In addition to the above approaches, 
\cite{goldenshluger2011bandwidth}
propose a method, known as the
Lepski's approach \citep{goldenshluger2008universal,lepski2009structural},
that treats the bandwidth selection problem as a model selection problem
and proposes a new criterion for selecting the smoothing bandwidth.
One feature of Lepski's approach is that the selected bandwidth
enjoys many statistical optimalities \citep{goldenshluger2011bandwidth}.

{\bf Remark.} (Kernel Selection)
In contrast to bandwidth selection, the choice of kernel function does not play
an important role in KDE.
The effect of the kernel function on the estimation error
is just a constant shift (via $\sigma_K$ and $\mu_K$ in equation \eqref{eq::PE}),
and the difference is generally very small among common kernel functions
(see, e.g., page 72 of \cite{wasserman2006all} and Section 6.2.3 in \cite{scott2015multivariate}),
so most of the literature ignore this topic.


\section{Confidence Intervals and Confidence Bands}	\label{sec::CS}

Confidence regions of the density function are random intervals
$C_{1-\alpha}(x)$ derived from the sample such that 
$C_{1-\alpha}(x)$
covers the true value of $p(x)$ with probability at least $1-\alpha$.
Based on this notion, there are two common types of confidence regions:
\begin{itemize}
\item \emph{Confidence interval}: for a given $x$, the set $C_{1-\alpha}(x)$ satisfies
$$
P\left(p(x)\in C_{1-\alpha}(x)\right) \geq 1-\alpha.
$$
\item \emph{Confidence band}: the interval $C_{1-\alpha}(x)$ satisfies
$$
P\left(p(x)\in C_{1-\alpha}(x)\,\,\forall x\in\K\right) \geq 1-\alpha.
$$
\end{itemize}
Namely, confidence intervals are confidence regions with only local coverage
and confidence bands are confidence regions with simultaneous coverage.
If a confidence interval/band has only coverage $1-\alpha+o(1)$, it will be 
called an \emph{asymptotically valid} $1-\alpha$ confidence interval/band.

For simplicity, we first ignore the bias between $\hat{p}_n(x)$ and $p(x)$
by assuming $p(x) = \mathbb{E}(\hat{p}_n(x))$
in Sections~\ref{sec::PCS} and \ref{sec::UCS} (i.e., we assume $B_h(x)=0$).
We will discuss strategies for handling the bias in Section \ref{sec::bias}

\subsection{Confidence Intervals}	\label{sec::PCS}

For a given point $x$,
by equation \eqref{eq::PE},
\begin{equation}
\sqrt{nh^d} \left(\hat{p}_n(x)-\mathbb{E}\left(\hat{p}_n(x)\right)\right) = \sqrt{nh^d} \cE_n \overset{d}{\rightarrow} N(0, \sigma^2_p(x)),
\label{eq::CLT}
\end{equation}
where $\sigma_p^2(x) = \mu_K\cdot p(x)$.
Equation \eqref{eq::CLT} implies that a straight-forward approach to construct a  
confidence band is to use asymptotic normality with a variance estimator.

{\bf Method 1: Plug-in Approach.}
A simple method is replacing $p(x)$ in the asymptotic variance by its estimator $\hat{p}_n(x)$,
leading to the following $1-\alpha$ confidence interval of $p(x)$:
\begin{equation}
C_{1-\alpha, \sf PI}(x) 
= \left[\hat{p}_n(x) - z_{1-\alpha/2}\sqrt{\frac{\mu_K\cdot\hat{p}_n(x)}{nh^d}},\quad
\hat{p}_n(x) + z_{1-\alpha/2}\sqrt{\frac{\mu_K\cdot\hat{p}_n(x)}{nh^d}}\right].
\label{eq::naive}
\end{equation}
We call this method the ``plug-in method" because we plug-in the variance estimator
to construct a confidence interval.
When $h\rightarrow 0, nh^d \rightarrow \infty$, $\hat{p}_n(x)$ is a consistent estimator of $p(x)$.
As a result, 
$$
P\left(\mathbb{E}(\hat{p}_n(x))\in C_{1-\alpha, \sf PI}(x)\right) = 1-\alpha +o(1).
$$

{\bf Method 2: Bootstrap and Plug-in Approach.}
An alternative method is to estimate the asymptotic variance 
using the bootstrap \citep{Efron1979}.
In more detail, we use the empirical bootstrap (also known as the nonparametric bootstrap or Efron's bootstrap, which
is to sample the original data with replacement) to generate bootstrap sample $X^*_1,\cdots, X^*_n$. 
Then, we apply KDE to the bootstrap sample, resulting in a bootstrap KDE $\hat{p}_n^*(x)$. 
When we repeat the bootstrap $B$ times, we then have $B$ bootstrap KDEs 
$\hat{p}_n^{*(1)}(x),\cdots, \hat{p}_n^{*(B)}(x)$. 
Let 
$$
\hat{\sigma}^2_{p, \sf BT}(x) = \frac{1}{B-1}\sum_{j=1}^B\left(\hat{p}_n^{*(j)}(x) - \bar{p}_n^{*}(x)\right)^2,
$$
where $\bar{p}_n^{*}(x) = \frac{1}{B}\sum_{j=1}^B \hat{p}_n^{*(j)}(x)$ is the sample average of the bootstrap KDE's.
Namely, $\hat{\sigma}^2_{p, \sf BT}(x)$ is the sample variance of the $B$ bootstrap KDE's evaluated at $x$.
A bootstrap $1-\alpha$ confidence interval is
\begin{equation}
C_{1-\alpha, \sf BT+PI}(x) 
= \left[\hat{p}_n(x) - z_{1-\alpha/2}\cdot\hat{\sigma}^2_{p, \sf BT}(x),\quad
\hat{p}_n(x) + z_{1-\alpha/2}\cdot\hat{\sigma}^2_{p, \sf BT}(x)\right].
\label{eq::BT_naive}
\end{equation}
Because the bootstrap variance estimator $\hat{\sigma}^2_{p, \sf BT}(x)$ converges to $\frac{\sigma^2_p(x)}{nh^d}$
in the sense that 
$$
\frac{\hat{\sigma}^2_{p, \sf BT}(x)}{\sigma^2_p(x)/(nh^d)}\overset{P}{\rightarrow}1,
$$
the bootstrap variance estimator is consistent, so the confidence interval will also be consistent:
$$
P\left(\mathbb{E}(\hat{p}_n(x))\in C_{1-\alpha, \sf BT+PI}(x)\right) = 1-\alpha +o(1).
$$

{\bf Method 3: Bootstrap Approach.}
In addition to the above methods, one can use a fully bootstrapping approach 
to construct a confidence interval without using asymptotic normality.
Let $\hat{p}_n^{*(1)}(x),\cdots, \hat{p}_n^{*(B)}(x)$ be bootstrap KDE's as in the previous method.
We define a pointwise deviation of a bootstrap KDE by
$$
\Delta_1(x) =  |\hat{p}_n^{*(1)}(x)- \hat{p}_n(x)|, \cdots, \Delta_B(x) =  |\hat{p}_n^{*(B)}(x)- \hat{p}_n(x)|.
$$
Then we compute the $1-\alpha$ quantile of the empirical CDF of $\Delta_1(x), \cdots, \Delta_B(x)$:
$$
c_{1-\alpha, \sf BT}(x) = \hat{G}_{x}^{-1}(1-\alpha), \quad \hat{G}_x(t) = \frac{1}{B} \sum_{j=1}^B I(\Delta_j \leq t).
$$
A $1-\alpha$ confidence interval of $p(x)$ is
\begin{equation}
C_{1-\alpha, \sf BT}(x) 
= \left[\hat{p}_n(x) - c_{1-\alpha, \sf BT}(x),\quad
\hat{p}_n(x) + c_{1-\alpha, \sf BT}(x)\right].
\label{eq::BT}
\end{equation}
Because the distribution of $|\hat{p}_n^{*}(x)- \hat{p}_n(x)|$ approximates 
the distribution of $|\hat{p}_n(x)- \mathbb{E}\left(\hat{p}_n(x)\right)|$, 
this confidence interval is also asymptotically valid, i.e.,
$$
P\left(\mathbb{E}(\hat{p}_n(x))\in C_{1-\alpha, \sf BT}(x)\right) = 1-\alpha +o(1).
$$

\begin{figure}
\center
\includegraphics[width=1.8in]{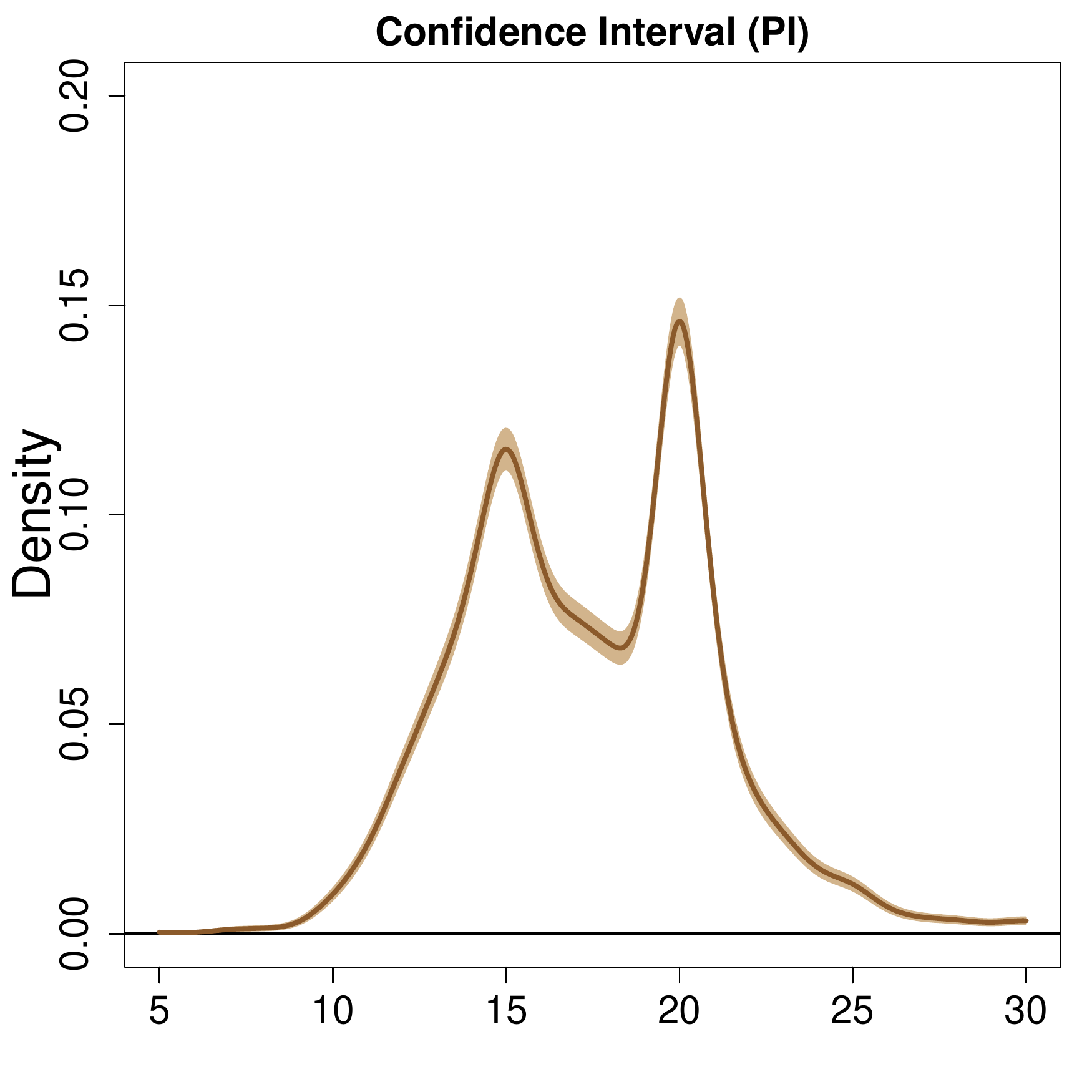}
\includegraphics[width=1.8in]{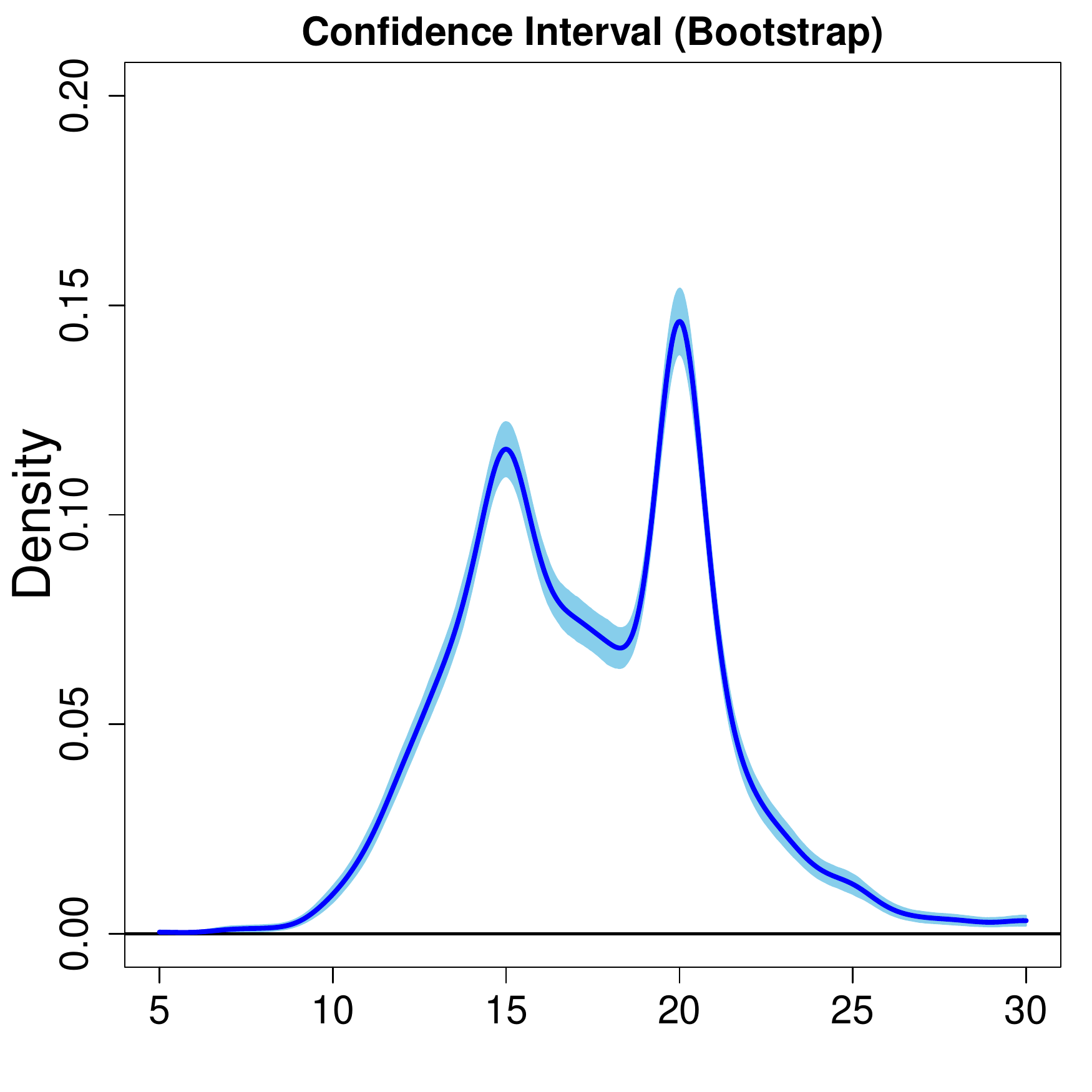}
\includegraphics[width=1.8in]{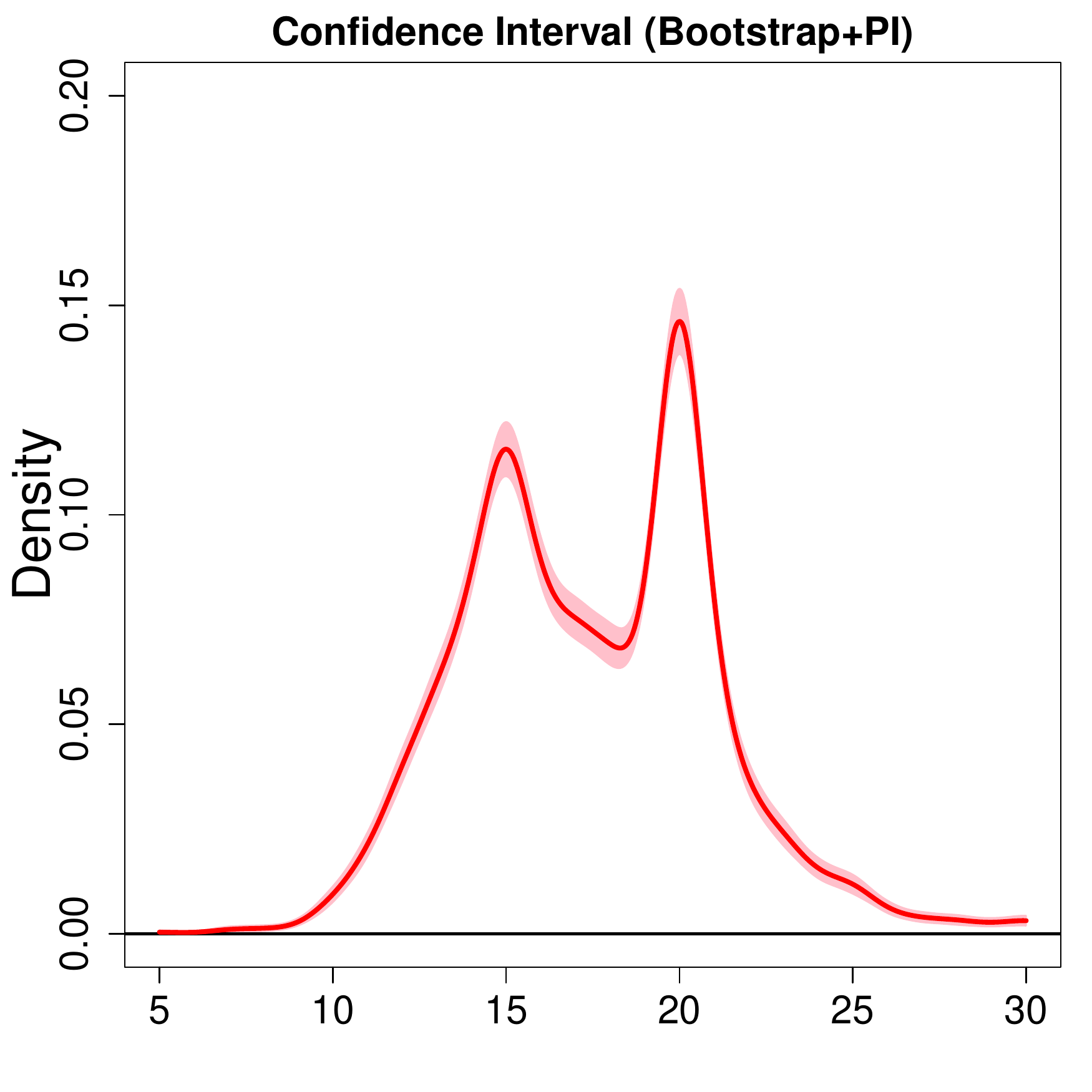}
\caption{$95\%$ confidence intervals from KDE.
We use the same data as in the left panel of
Figure~\ref{fig::ex0}.
{\bf Left:} 
We obtain confidence intervals using the plug-in approach (method 1 in Section~\ref{sec::PCS}).
{\bf Middle:} 
We construct confidence intervals using the plug-in approach with the bootstrap (method 2 in Section~\ref{sec::PCS}).
{\bf Right:} 
We build confidence intervals using the bootstrap approach (method 3 in Section~\ref{sec::PCS}).
The three confidence regions are nearly the same, although
they are constructed using different approaches.
Note that there are two problems with these confidence regions.
First, since we ignore the bias, the actual coverage might be substantially less than the nominal coverage $95\%$.
Second, because they are confidence intervals, the we only have pointwise coverage.
Thus, even though the actual coverage is guaranteed, 
these regions might not cover the entire actual density function. 
}
\label{fig::CS1}
\end{figure}

\subsubsection{Example: Confidence Intervals}

In Figure~\ref{fig::CS1}, we compare the three approaches of constructing a confidence interval
using the NACC Uniform Data Set, as described in the Introduction (Section~\ref{sec::intro}) and the left panel of Figure~\ref{fig::ex0}.
The left panel is the 95\% confidence interval of each point using the plug-in approach (method 1);
the middle panel is the 95\% confidence interval from the plug-in and bootstrap approach (method 2);
the right panel is the 95\% confidence interval from the bootstrap approach (method 3).

Essentially, the three confidence intervals are very similar;
in particular, the confidence intervals from the method 2 and 3 are nearly identical.
The interval from method 1 is slightly smaller than the other two.

While all of these intervals are valid for each given point, there is no guarantee that
they will cover the entire density function \emph{simultaneously}. 
In the next section, we introduce methods of constructing confidence bands (confidence regions
with simultaneous coverage).

\subsection{Confidence Bands}		\label{sec::UCS}

Now, we present methods of constructing confidence bands.
The key idea is to approximate the distribution of the uniform error $\sup_x |\hat{p}_n(x)-p(x)|$
and then convert it into a confidence band.
To be more specific, let $G(t) = P(\sup_x |\hat{p}_n(x)-p(x)|<t)$ be the CDF of the uniform
error, and let $\bar{c}_{1-\alpha} = G^{-1}(1-\alpha)$ be the $1-\alpha$ quantile.
Then it can be shown that the set
$$
\bar{C}(x) = \left[\hat{p}_n(x)-\bar{c}_{1-\alpha},\,\, \hat{p}_n(x)+\bar{c}_{1-\alpha}\right]
$$
is a confidence band, i.e., 
$$
P\left(p(x) \in \bar{C}(x) \,\,\forall x\in\K\right) = 1-\alpha.
$$
Therefore, 
as long as we have a good approximation of the distribution $G(t)$,
we can convert the approximation into a confidence band.


{\bf Method 1: Plug-in Approach.}
An intuitive approach is to derive the asymptotic distribution of $\sup_x |\hat{p}_n(x)-p(x)|$
directly and then invert it 
into a confidence band.
\cite{bickel1973some, rosenblatt1976maximal} proved that the uniform loss converges to an extreme value distribution
in the sense that
\begin{equation}
P\left(\sqrt{-2\log h}\left(\sqrt{nh^d}\sup_x \frac{|\hat{p}_n(x)-\mathbb{E}(\hat{p}_n(x))|}{\sqrt{p(x)\mu_K}}-d_n\right)<t\right) \rightarrow e^{-2e^{-t}},
\label{eq::EVT}
\end{equation}
where $d_n=O(\sqrt{-2\log h})$ is a quantity depending only on $n,h$ and the kernel function $K$.
\cite{rosenblatt1976maximal} provided an exact expression for the quantity $d_n$.
Let $E_{1-\alpha} = -\log\left(-\frac{\log\alpha}{2}\right)$ be the $1-\alpha$ quantile
of the right-hand-side CDF.
Define
$$
c_{1-\alpha} =\sqrt{\frac{p(x)\mu_K}{nh^d}} \left(d_n +\frac{E_{1-\alpha}}{\sqrt{-2\log h}}\right).
$$
Then, by equation \eqref{eq::EVT},
$\sup_x |\hat{p}_n(x)-\mathbb{E}(\hat{p}_n(x))|$ falls within $[0,c_{1-\alpha}]$
with probability at least (asymptotically) $1-\alpha$.
To construct a confidence band, we replace the quantity $p(x)$ in $c_{1-\alpha}$ with a plug-in estimate from KDE,
leading to
$$
c_{1-\alpha, \sf PI} = \sqrt{\frac{\hat{p}_n(x)\mu_K}{nh^d}} \left(d_n +\frac{E_{1-\alpha}}{\sqrt{-2\log h}}\right).
$$
Then a $1-\alpha$ confidence band will be 
\begin{equation}
C^\dagger_{1-\alpha, \sf PI}(x)
= \left[\hat{p}_n(x) - c_{1-\alpha, \sf PI},\quad
\hat{p}_n(x) + c_{1-\alpha, \sf PI}\right].
\label{eq::PI_uni}
\end{equation}

Although equation \eqref{eq::PI_uni} is an asymptotically valid confidence band,
the convergence to the extreme value distribution in equation \eqref{eq::EVT} is very slow \cite{hall1991convergence}.
Thus, we need a huge sample size to guarantee that the confidence band from \eqref{eq::PI_uni} is asymptotically valid.
To resolve this problem, we use the bootstrap. 

{\bf Method 2: Bootstrap Approach.}
The key element of how the bootstrap works is that the uniform error can be approximated accurately by the supremum of
a Gaussian process \citep{neumann1998strong,chernozhukov2014comparison,chernozhukov2014gaussian}.
In more detail, there exists a tight Gaussian process $B_n(x)$ such that
\begin{equation}
\sup_t \left|P\left(\sqrt{nh^d}\sup_x\left|\hat{p}_n(x) - \mathbb{E}(\hat{p}_n(x))\right|<t \right)- 
P\left(\sup_x\left|B_n(x)\right|<t \right)\right| = o(1).
\label{eq::Gaussian}
\end{equation}
Moreover, the difference between the bootstrap KDE and the original KDE also has a similar convergent result
\citep{chernozhukov2013gaussian,chernozhukov2014anti,chernozhukov2016empirical}:
\begin{equation}
\sup_t \left|P\left(\sqrt{nh^d}\sup_x\left|\hat{p}^*_n(x) - \hat{p}_n(x)\right|<t \big|X_1,\cdots,X_n\right)- 
P\left(\sup_x\left|B_n(x)\right|<t \right)\right| = o_P(1),
\label{eq::BT_Gaussian}
\end{equation}
where $B_n(x)$ is the same Gaussian process as the one in equation \eqref{eq::Gaussian}.
Thus, the distribution of $\sup_x\left|\hat{p}_n(x) - \mathbb{E}(\hat{p}_n(x))\right|$
will be approximated by the distribution of its bootstrap version $\sup_x\left|\hat{p}^*_n(x) - \hat{p}_n(x)\right|$.
As a result, the bootstrap quantile of uniform error converges to the quantile of the actual uniform error,
thereby proving that the bootstrap confidence band is asymptotically valid.

Here is the formal construction of a bootstrap confidence band.
Let $\hat{p}_n^{*(1)}(x),\cdots, \hat{p}_n^{*(B)}(x)$ be the bootstrap KDE's.
We define the uniform deviation of the bootstrap KDE's by
$$
\Delta_1 =  \sup_x|\hat{p}_n^{*(1)}(x)- \hat{p}_n(x)|, \cdots, \Delta_B =  \sup_x|\hat{p}_n^{*(B)}(x)- \hat{p}_n(x)|.
$$
Then, we compute the $1-\alpha$ quantile of the empirical CDF of $\Delta_1, \cdots, \Delta_B$ as
$$
c_{1-\alpha, \sf BT} = \hat{G}_{\K}^{-1}(1-\alpha), \quad \hat{G}_{\K}(t) = \frac{1}{B} \sum_{j=1}^B I(\Delta_j(x) \leq t).
$$
A $1-\alpha$ confidence band will be
\begin{equation}
C^\dagger_{1-\alpha, \sf BT}(x)
= \left[\hat{p}_n(x) - c_{1-\alpha, \sf BT},\quad
\hat{p}_n(x) + c_{1-\alpha, \sf BT}\right].
\label{eq::BT_uni}
\end{equation}
By equations \eqref{eq::Gaussian} and \eqref{eq::BT_Gaussian},
when $B\rightarrow \infty$,
$$
P\left(\mathbb{E}(\hat{p}_n(x))\in C^\dagger_{1-\alpha, \sf BT}(x)\,\,\forall x\in \K\right) = 1-\alpha +o(1).
$$
Namely, the set $C^\dagger_{1-\alpha, \sf BT}(x)$ is an asymptotically valid $1-\alpha$ confidence band.

\begin{figure}
\center
\includegraphics[width=2in]{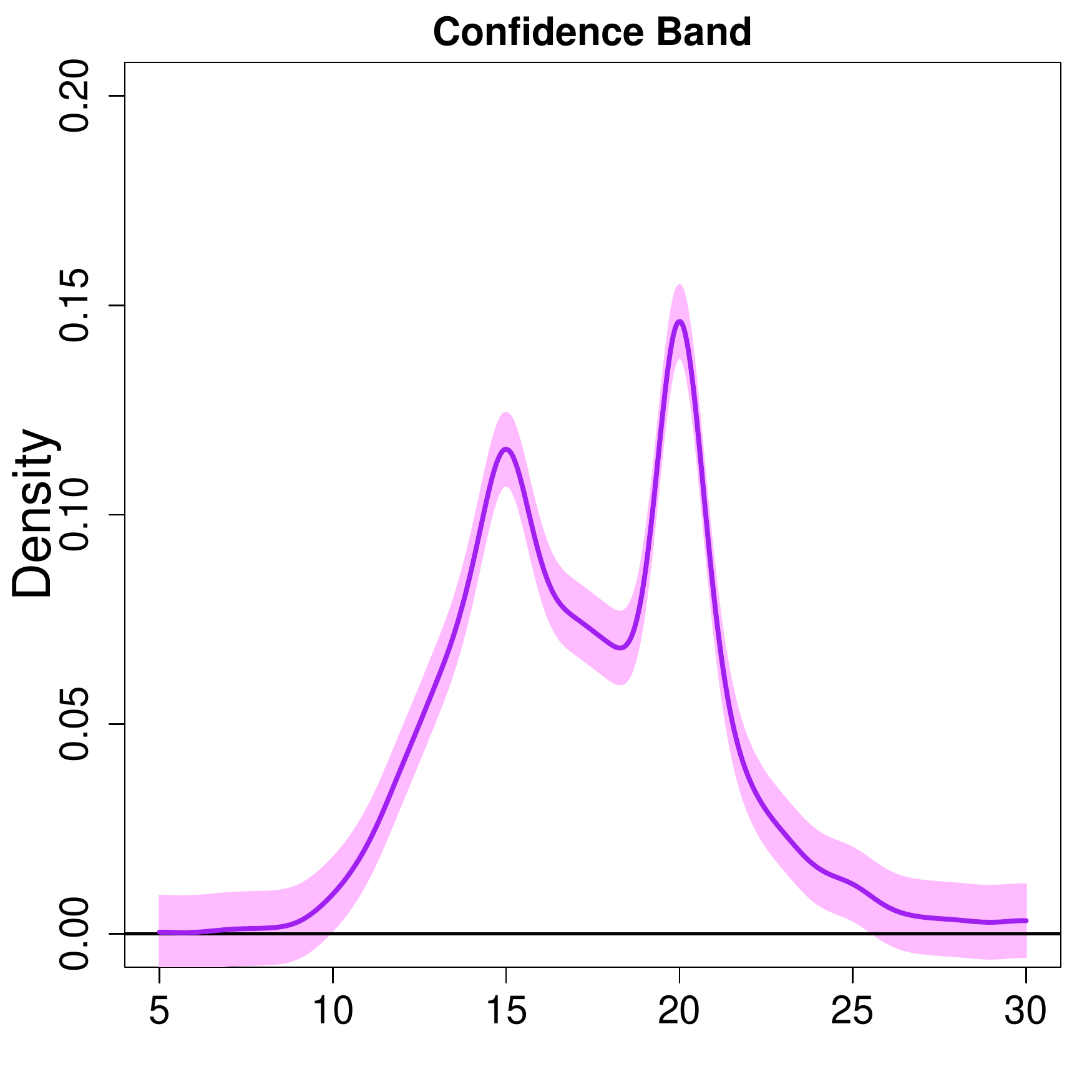}
\includegraphics[width=2in]{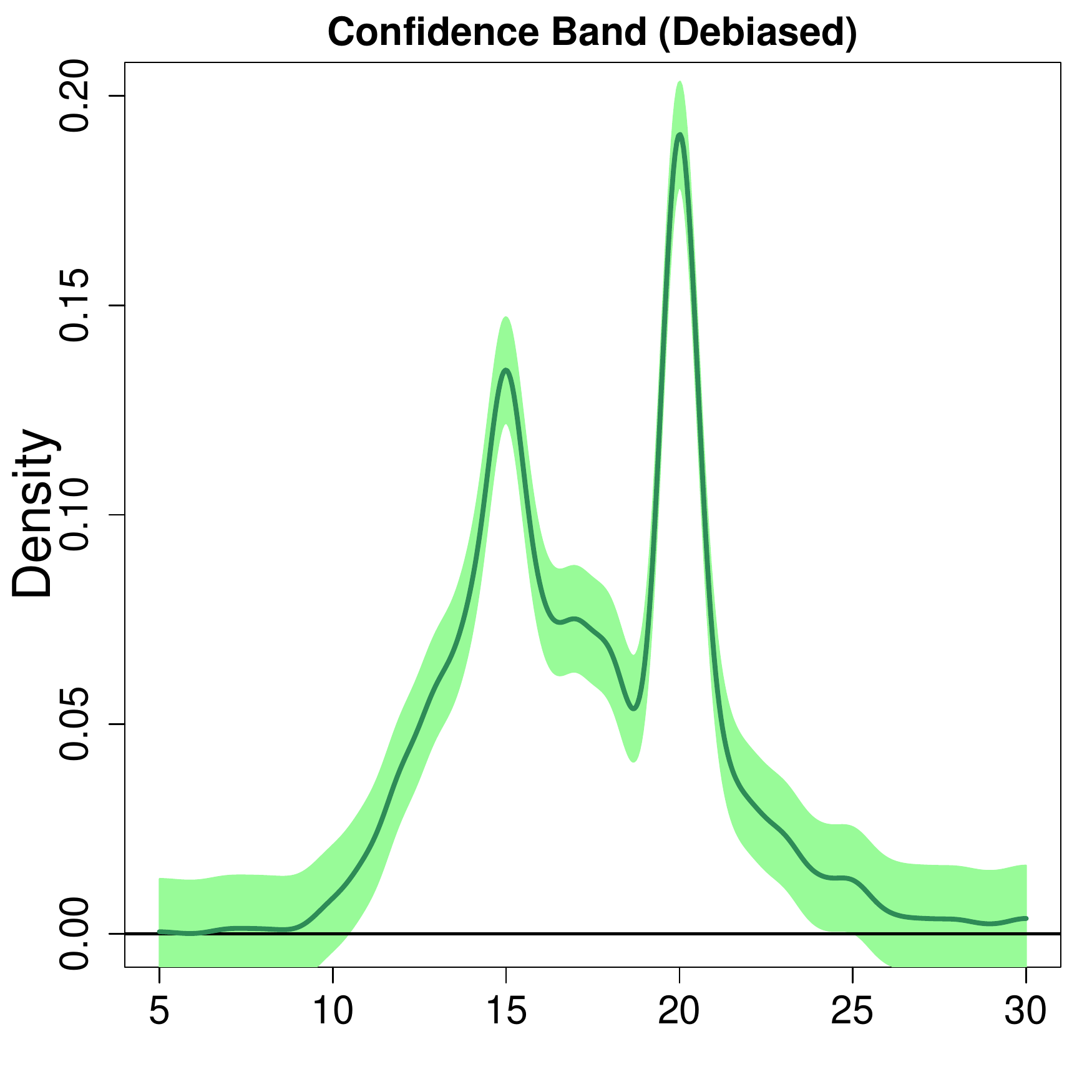}
\caption{$95\%$ confidence bands from KDE via the bootstrap.
This is the same dataset as in the left panel of Figure~\ref{fig::ex0}.
{\bf Left:} 
The confidence band from bootstrapping the uniform error of KDE (method 2 of Section~\ref{sec::UCS}).
Although this is a confidence band,
we ignore the bias in constructing the confidence band so the actual coverage
could be below the nominal coverage. 
{\bf Right:} 
The confidence band from bootstrapping the uniform error of the debiased KDE
(method proposed in Figure~\ref{fig::alg::KDE}).
Because we are using the debiased KDE, the density curve is slightly different
from the blue curve in the left panel. 
Although the confidence band is wider than the confidence in the left,
this confidence band has coverage guaranteed 
when the smoothing bandwidth is chosen at rate $O\left(n^{-\frac{1}{d+4}}\right)$.
Note that there are possible approach to narrowing down the size of this confidence band;
we refer the reader to \cite{chen2017nonparametric}. 
}
\label{fig::CS2}
\end{figure}

\subsubsection{Example: Confidence Bands}

Figure~\ref{fig::CS2} presents confidence bands using KDE. 
The left panel shows the confidence band from the bootstrap approach introduced in the previous section.
Compared to the confidence intervals in Figure \ref{fig::CS1}, the confidence bands are wider
because we need to control the coverage simultaneously for every point. 

However, the confidence band in the left panel of Figure~\ref{fig::CS2} 
(and all of the confidence intervals in Figure~\ref{fig::CS1})
has a serious problem--the coverage guarantee is for the expected value of KDE $\E\left(\hat{p}_n(x)\right)$
rather than for the true density function $p$.
Unless we undersmooth KDE (choose $h$ converging at a faster rate than $O(n^{-\frac{1}{d+4}})$), 
the confidence band shows undercoverage. 
We will discuss this topic in more detail in Section~\ref{sec::bias}.

To construct an (asymptotically valid) confidence band with $h$ being at rate $O(n^{-\frac{1}{d+4}})$, 
we use the debiased estimator introduced in \cite{chen2017nonparametric}.
The details are provided in Section~\ref{sec::BC} and Figure~\ref{fig::alg::KDE}.
The right panel of Figure~\ref{fig::CS2} shows a confidence band from the debiased KDE approach. 
Although the confidence band is wider, the coverage is guaranteed for such a confidence band.


\subsection{Handling the Bias}	\label{sec::bias}

In the previous section, we ignored the bias in KDE.
However, the bias could be a severe problem in reality because it systematically
shifted our confidence interval/band so the actual coverage is below the nominal coverage.
Here we discuss 
strategies to handle bias.

\subsubsection{Ignoring the Bias}
A simple strategy is to ignore the bias and focus on inferring
the expectation of KDE $p_h(x) = \E\left(\hat{p}_n(x)\right)$.  
$p_h$ is called 
the smoothed or mollified density function \citep{Rinaldo2010a,chen2015asymptotic,chen2015density}. 
As long as the kernel function $K$ is smooth,
$p_h$ will also be smooth.
Moreover, $p_h$ exists even when the distribution function is singular (in this case, the population density $p$ does not exist). 
For inferring geometric or topological features, 
$p_h$ might be a better parameter of interest
because structures in $p_h$ generally represent salient structures of $p$ \citep{chen2015density}
and many topological structures of $p_h$ will be similar to those of $p$ when $h$ is small 
\citep{chen2015asymptotic,genovese2014nonparametric,jisu2016statistical}.
If we switch our target to $p_h$, we have to
make it clear that this is a confidence region of $p_h$ rather than of $p$
when we report our confidence regions.


\subsubsection{Undersmoothing}

Undersmoothing is a very common approach to handle bias in KDE (and other nonparametric approaches).
Recall from equation \eqref{eq::PE}:
\begin{align*}
\hat{p}_n(x) - p (x) &= B_h(x) = \cE_n(x)\\
& = O(h^2) + O_P\left(\sqrt{\frac{1}{nh^d}}\right).
\end{align*}
The bias is $B_h(x) = O(h^2)$, and the stochastic variation is the $\cE_n(x) = O_P\left(\sqrt{\frac{1}{nh^d}}\right)$ term.
If now we take $h\rightarrow 0$ such that 
$h^2 =  o\left(\sqrt{\frac{1}{nh^d}}\right)$, then
the stochastic variability dominates the errors. 
Therefore, we can ignore the bias term
and use the method suggested in Section~\ref{sec::PCS} and \ref{sec::UCS}.

Note that $h^2 =  o\left(\sqrt{\frac{1}{nh^d}}\right)$ is equivalent to $nh^{d+4}\rightarrow 0$,
which corresponds to choosing a smaller smoothing bandwidth than the optimal smoothing bandwidth ($h_{\sf opt} \asymp n^{-\frac{1}{d+4}}$)
that balances the bias and the variance (this is why it is called undersmoothing).
Although the undersmoothing provides a valid construction of confidence regions,
such a choice of bandwidth implies that 
the size of the confidence band is larger than the optimal size because we are inflating
the variance to eliminate the bias.
Some references to undersmoothing can be found in 
\cite{bjerve1985uniform, hall1992bootstrap, hall1993empirical, chen1996empirical, neumann1998simultaneous, chen2002confidence, mcmurry2008bootstrap}.

\subsubsection{Bias-corrected and Oversmoothing}	\label{sec::BC}

An alternative approach to construct a valid confidence band is to correct the bias of KDE explicitly;
this approach is known as the bias-corrected method and the resulting KDE is called the bias-corrected KDE.
Recall from equation \eqref{eq::PE} that the bias in KDE is
$$
\E\left(\hat{p}_n(x)\right) -p(x) = \frac{h^2}{2} \sigma_K^2 \nabla^2 p(x) +o(h^2).
$$
Thus, we can correct the bias by estimating $\nabla^2 p(x)$.
The quantity $\nabla^2 p(x)$ can be estimated by $\nabla^2 \hat{p}_b(x)$,
where 
$$
\hat{p}_b(x) = \frac{1}{nb^d}\sum_{i=1}^n K\left(\frac{x-X_i}{b}\right)
$$
is KDE using smoothing bandwidth $b$.
Recall from Section \ref{sec::deriv} that
the second derivative estimator has an error rate
\begin{equation}
\nabla^2 \hat{p}_b(x) - \nabla^2 p(x) = O(b^2) +O_P\left(\sqrt{\frac{1}{nb^{d+4}}}\right).
\label{eq::bKDE}
\end{equation}
Thus, to obtain a consistent estimator of $\nabla^2 p(x)$, 
we have to choose another smoothing bandwidth $b$, and this smoothing bandwidth needs to be 
larger than the optimal bandwidth $h_{\sf opt} = O(n^{-\frac{1}{d+4}})$.
Because the choice of $b$ corresponds to oversmoothing KDE,
this approach is also called the ``oversmoothing" method. 

Using $\hat{p}_b(x)$, the bias-corrected KDE
is
$$
\widetilde{p}_n(x) = \hat{p}_n(x) - \frac{h^2}{2}\sigma_K^2 \nabla^2 \hat{p}_b(x).
$$
When $\nabla^2 \hat{p}_b(x)$ is a consistent estimator of $\nabla^2 p(x)$,
the pointwise error rate is 
$$
\widetilde{p}_n(x) - p(x) =o(h^2)+o_P\left(h^2\right)+O_P\left(\sqrt{\frac{1}{nh^d}}\right),
$$
so the dominating quantity, $O_P\left(\sqrt{\frac{1}{nh^d}}\right)$, is the stochastic variation in $\widetilde{p}_n(x)$.
Thus,
the confidence regions can be constructed by replacing $\hat{p}_n(x)$ by $\widetilde{p}_n(x)$
in equations \eqref{eq::naive}, \eqref{eq::BT_naive}, \eqref{eq::BT}, and \eqref{eq::BT_uni}.
An incomplete list of literature of the bias-corrected approach is as follows:
\cite{hardle1988bootstrapping,hardle1991bootstrap,hall1992effect, eubank1993confidence, sun1994simultaneous, hardle1995better, neumann1995automatic, xia1998bias, hardle2004bootstrap}.

\cite{calonico2015effect} proposes a plug-in method of constructing a confidence interval with $b=h$.
Although the choice $b=h$ does not lead to a consistent estimate of the second derivative, 
the bias will be pushed into the next order because the bias-corrected estimator can be viewed as 
a higher order kernel function (see page 157 in \cite{scott2015multivariate}). 
Thus, since the dominating term in the estimation error is the stochastic variation,
we can construct an asymptotically valid confidence interval using the plug-in approach. 

\begin{figure}[h]
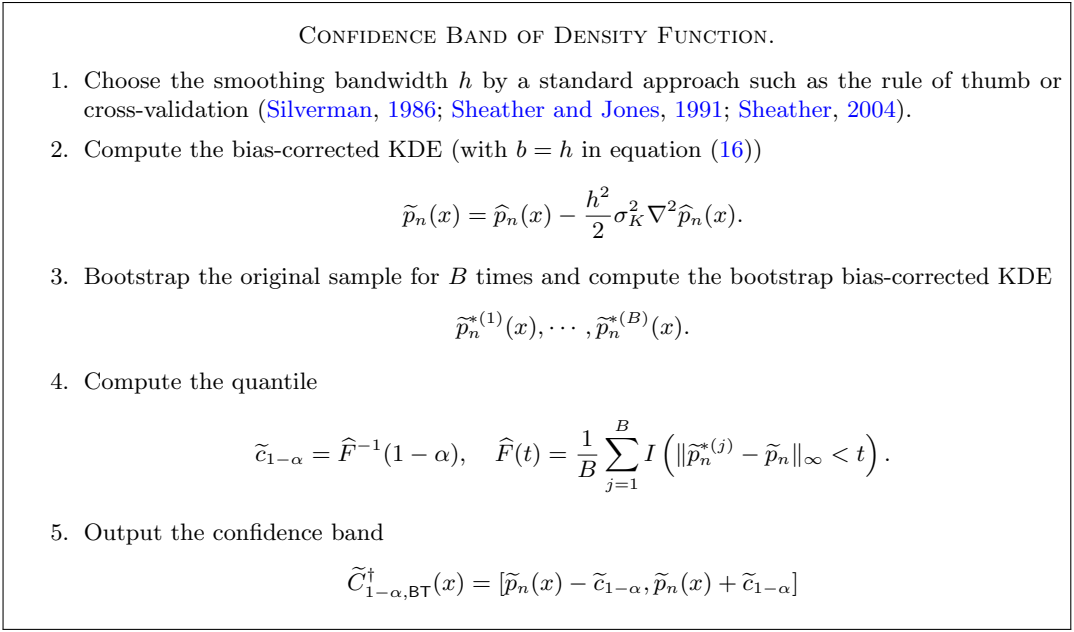

\fbox{\parbox{5.5in}{
\begin{center}
{\sc Confidence Band of Density Function.}
\end{center}
\begin{center}
\begin{enumerate}
\item 
Choose the smoothing bandwidth $h$ by a standard approach such as the rule of thumb or cross-validation \citep{Silverman1986,sheather1991reliable,sheather2004density}. 
\item 
Compute the bias-corrected KDE (with $b=h$ in equation \eqref{eq::bKDE})
$$
\widetilde{p}_n(x) = \hat{p}_n(x) - \frac{h^2}{2}\sigma_K^2 \nabla^2 \hat{p}_n(x).
$$
\item
Bootstrap the original sample for $B$ times and compute the bootstrap bias-corrected KDE
$$
\widetilde{p}_n^{*(1)}(x), \cdots, \widetilde{p}_n^{*(B)}(x).
$$
\item
Compute the quantile 
$$
\tilde{c}_{1-\alpha} = \hat{F}^{-1}(1-\alpha), \quad \hat{F}(t) = \frac{1}{B}\sum_{j=1}^BI\left(\|\widetilde{p}_n^{*(j)}-\widetilde{p}_n\|_\infty<t\right).
$$
\item 
Output the confidence band 
$$
\tilde{C}^\dagger_{1-\alpha, \sf BT}(x) = [\widetilde{p}_n(x)- \tilde{c}_{1-\alpha}, \widetilde{p}_n(x)+\tilde{c}_{1-\alpha}]
$$
\end{enumerate}
\end{center}
}}
\caption{A confidence band of the density function from bootstrapping the debiased KDE \citep{chen2017nonparametric}.
This confidence band is asymptotically valid and is compatible with most of the bandwidth selectors
introduced in Section~\ref{sec::BS}.}
\label{fig::alg::KDE}
\end{figure}

\cite{chen2017nonparametric} further generalizes the idea of \cite{calonico2015effect}
to confidence bands by bootstrapping the bias-corrected kernel density estimator with $b=h$. 
Figure~\ref{fig::alg::KDE} summarizes the procedure of constructing a confidence band using the approach in \cite{chen2017nonparametric}.
The resulting confidence band, $\tilde{C}^\dagger_{1-\alpha, \sf BT}(x)$, has the following property:
$$
P\left(p(x)\in\tilde{C}^\dagger_{1-\alpha, \sf BT}(x) \,\, \forall x\in\K\right) = 1-\alpha+o(1)
$$
when $h=h_{\sf opt}=O(n^{-\frac{1}{d+4}})$.
Namely, $\tilde{C}^\dagger_{1-\alpha, \sf BT}(x)$ is an asymptotically valid $1-\alpha$ confidence band
when we pick $h$ under the optimal rate.
The right panel of Figure~\ref{fig::CS2} shows an example
of this confidence band. 
Although this approach generally leads to a wider confidence band, 
this confidence band has asymptotically $1-\alpha$ coverage whereas the confidence band
in the left panel of Figure~\ref{fig::CS2} has undercoverage.

{\bf Remark.} (Calibration)
In addition to the above methods, 
another possible approach is to choose a corrected coverage of confidence regions; this 
approach is called ``calibration" and is related to the work in
\cite{beran1987prepivoting, hall1986bootstrap, loh1987calibrating, hall2013simple}.
The principal idea is to investigate the effect of the bias on the coverage
of the confidence band
and then choose a conservation quantile to guarantee the nominal coverage
of the resulting confidence regions.

\section{Geometric and Topological Features}	\label{sec::GT}

KDE can be used to estimate not only the underlying density function but also
geometric (and topological) structures related to the density.
To be more precise, 
many geometric (and topological) features of $\hat{p}_n$
converges to the corresponding structures of $p$,
and hence, we can use a structure of $\hat{p}_n$ as the estimator 
of that structure of $p$.

Because geometric and topological structures generally involve
the gradient and Hessian matrix of the density function,
we define some notations here.
We define 
$g(x) = \nabla p(x)$ to be the gradient of the density
and $H(x) = \nabla \nabla p(x)$ to be the Hessian matrix of the density.
Moreover, we also define $\lambda_1(x)\geq \cdots\geq \lambda_d(x)$
to be the largest to the smallest eigenvalues of $H(x)$
and $v_1(x),\cdots, v_d(x)$ to be the corresponding eigenvectors.

\begin{figure}
\center
\includegraphics[width=2in]{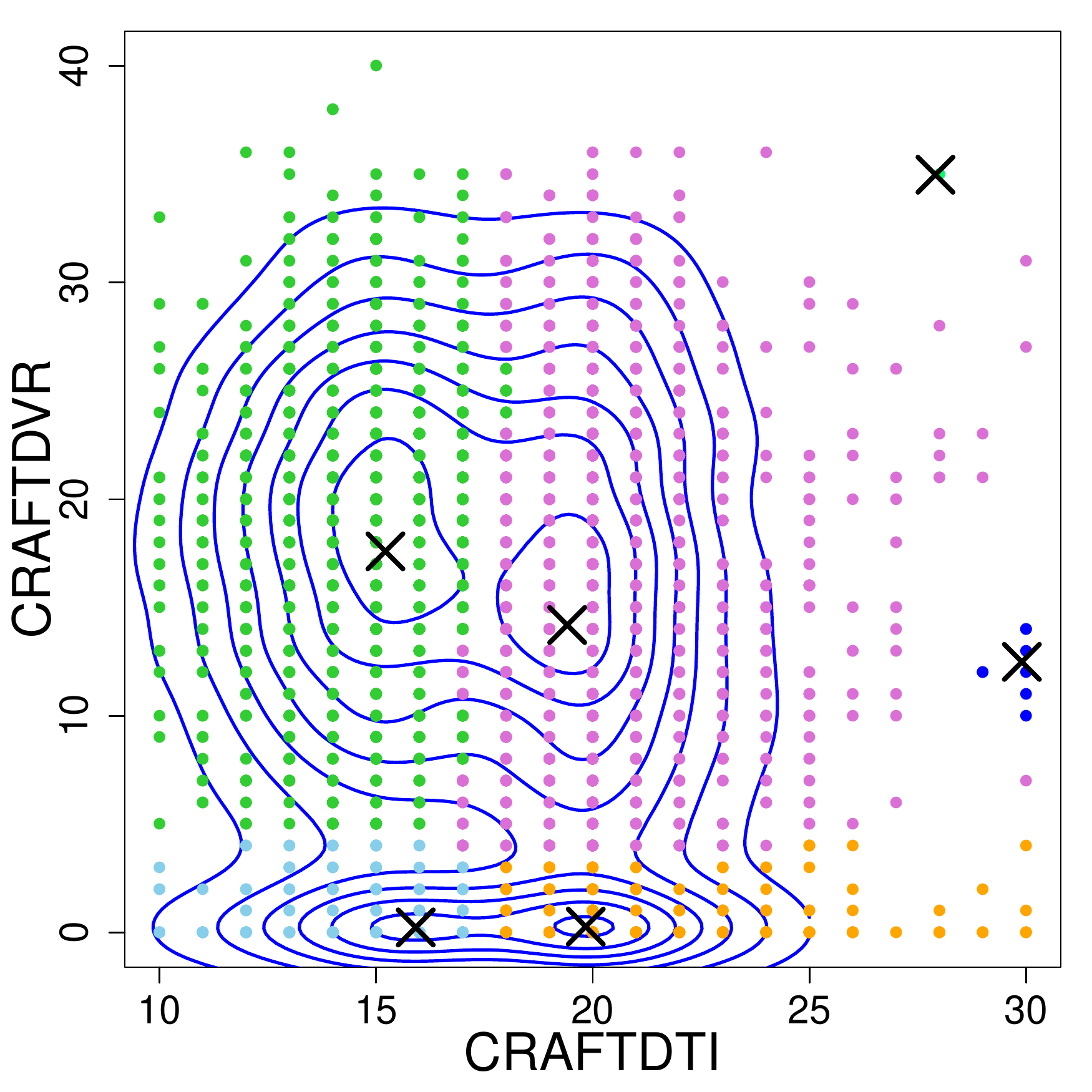}
\includegraphics[width=2in]{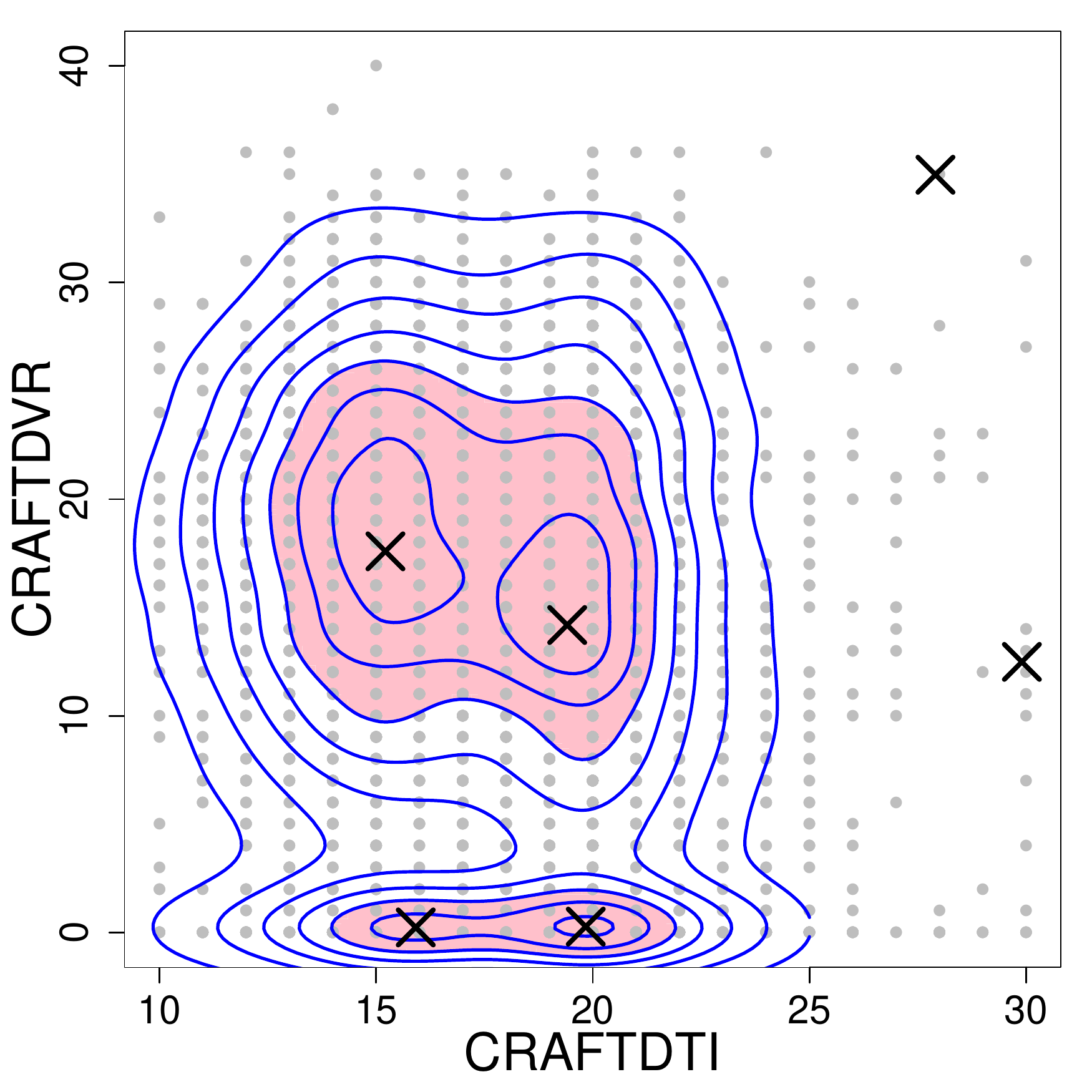}
\caption{
Estimating geometric features using KDE.
This is the same dataset as the right panel of Figure~\ref{fig::ex0}.
{\bf Left:} 
Density local modes (black crosses) and mode clustering.
The colored points describe the clusters that the respective subject belong to.
Mode clustering uses the gradient flow to partition data points into clusters.
Thus, each cluster (points with the same color) has a local mode as its representative. 
{\bf Right:} 
Density contours (blue), local modes (black crosses), and a density level set (pink area).
A density level set is just a region containing points whose density values are greater than or equal to a
particular level. 
Thus, it will contain regions within some contour lines (blue curves). 
}
\label{fig::geo1}
\end{figure}

\subsection{Local Modes}

A well-known geometric feature of the density function is its (global) mode.
Actually, when Parzen introduced KDE,
he mentioned the use of the mode of KDE to estimate the mode of the density function \citep{parzen1962estimation}.
The asymptotic distribution and confidence sets of the mode were later discussed in 
\cite{romano1988bootstrapping,romano1988weak}.

We can extend the definition of the (global) mode to a local sense and define the local modes:
$$
\mathcal{M} = \{x: g(x) = 0, \lambda_1(x)<0\}.
$$
Namely, $\mathcal{M}$ is the collection of points for which the density function is locally maximized. 
A natural estimator of $\mathcal{M}$ is a plug-in from KDE \citep{chazal2014robust,chen2016comprehensive}:
$$
\hat{\mathcal{M}} = \{x: \hat{g}_n(x) = 0, \hat{\lambda}_1(x)<0\},
$$
where $\hat{g}_n(x)$ and $\hat{\lambda}_1(x)$ are KDE version of $g(x)$ and $\lambda_1(x)$.
Under mild assumptions, $\hat{\mathcal{M}}$ is a consistent estimator of $\mathcal{M}$
\citep{chazal2014robust,chen2016comprehensive}.
Note that one can use the mean shift algorithm \citep{fukunaga1975estimation,cheng1995mean,comaniciu2002mean} to compute
the estimator $\hat{\mathcal{M}}$ numerically.

Note that one can use the local modes to cluster data points;
this 
is called mode clustering \citep{chacon2015population,azizyan2015risk,chen2016comprehensive} 
or mean-shift clustering \citep{fukunaga1975estimation,cheng1995mean,comaniciu2002mean}. 
The left panel of Figure~\ref{fig::geo1}
shows a case of estimated local modes (black boxes) and mode clustering
using the mean-shift algorithm. 
In R, one can use the library `{\sf LPCM}\footnote{\url{https://cran.r-project.org/web/packages/LPCM/index.html}}'
to compute the estimator $\hat{\mathcal{M}}$ and perform mode clustering.


\subsection{Level Sets}

Level sets are regions for which the density value is equal to or above a particular level.
Given a level $\lambda$, the $\lambda$-level set \citep{Polonik1995, Tsybakov1997} is
$$
L_\lambda = \{x: p(x)\geq \lambda\}. 
$$
A natural estimator of $L_\lambda$ is the plug-in estimate from KDE:
$$
\hat{L}_\lambda = \{x: \hat{p}_n(x)\geq \lambda\}. 
$$
The pink area in the right panel of Figure~\ref{fig::geo1}
is one instance of the density level set.

There is substantial statistical literature discussing different types of
convergence of $\hat{L}_\lambda$;
see \cite{Polonik1995, Tsybakov1997,Rinaldo2010a,Rinaldo2010b} and the references therein. 
\cite{mammen2013confidence} and \cite{chen2015density} propose procedures
for constructing confidence sets of $L_\lambda$ through bootstrapping the estimator $\hat{L}_\lambda$. 
Note that a visualization tool\footnote{R source code: \url{https://github.com/yenchic/HDLV}} for multivariate level sets is proposed in 
\cite{chen2015density}.

\subsection{Ridges}

Another interesting geometric structures are ridges
\citep{genovese2014nonparametric,chen2015asymptotic,qiao2016theoretical} of the density functions.
Formally, ridges are defined as follows.
Let $V(x) = [v_2(x) \cdots v_d(x)]\in \R^{d\times (d-1)}$ be the matrix consisting of the second eigenvector
to the last eigenvector. 
A density ridge is defined as
$$
\cR = \{x: V(x)V(x)^Tg(x) = 0, \lambda_2(x)<0\}. 
$$
Intuitively, any point $x\in \cR$ is a local mode in the subspace spanned by $v_2(x),\cdots,v_d(x)$.
Thus, if we move away from $\cR$ in the subspace, the density value decreases, which is the characteristic attribute of a ridge. 

To estimate $\cR$, we again use the plug-in from KDE:
$$
\hat{\cR} = \left\{x: \hat{V}(x)\hat{V}(x)^T\hat{g}(x) = 0, \hat{\lambda}_2(x)<0\right\},
$$
where $\hat{V}(x)$ and $\hat{\lambda}_2(x)$ are KDE versions of $V(x)$ and $\lambda_2(x)$ respectively.
The convergence rate and topological characteristics were discussed in \cite{genovese2014nonparametric}.
\cite{chen2015asymptotic} and \cite{qiao2016theoretical} both studied
the asymptotic theory, and \cite{chen2015asymptotic} further proposed methods of constructing confidence sets of $\cR$.
\cite{Ozertem2011} introduced the subspace-constrained mean shift (SCMS) algorithm to compute $\hat{\cR}$.
The red curves in the right panel of Figure~\ref{fig::geo1}
are estimated ridges from the SCMS algorithm.

%
%
%

\subsection{Morse-Smale Complex}

The Morse-Smale complex \citep{banyaga2004lectures} of a density function $p$ is a partition of the entire support $\K$
based on the density gradient flow.
For any point $x\in\K$, we define a gradient ascent flow $\pi_x(t)$ such that
$$
\pi_x'(t) = g(\pi_x(t)), \quad\pi_x(0) = x.
$$
Namely, $\pi_x(t)$ is a flow starting at $x$ such that we move along the orientation of the density gradient ascent. 
By Morse theory \citep{morse1925relations,morse1930foundations}, 
such a flow converges to a destination that is one of the critical points 
(points where $g(x)=0$) when the density function is smooth.

Similarly, we define a gradient descent flow $\gamma_x(t)$ such that
$$
\gamma_x'(t) = -g(\gamma_x(t)), \quad\gamma_x(0) = x.
$$
In a similar manner to $\pi_x(t)$, $\gamma_x(t)$ starts at $x$ but now the flow moves by following density gradient descent. 
Again, by Morse theory, such a flow converges also to one of the critical points, but not to the same point
as the destination of $\pi_x(t)$. 
Based on the destinations of $\pi_x(t)$ and $\gamma_x(t)$, we partition the entire support $\K$
into different regions; points within the same region
share the same destination for both the gradient ascent flow and the gradient descent flow.
This partition is called the Morse-Smale complex.

To estimate the Morse-Smale complex of $p$,
we use the Morse-Smale complex of $\hat{p}_n$. 
\cite{arias2016estimation} and \cite{chen2015statistical} studied
the convergence of gradient flows and the Morse-Smale complex
of $\hat{p}_n$ and proved the statistical consistency of these geometric features.
\cite{chen2015statistical} further proposed to use the Morse-Smale complex to
visualize a multivariate density function\footnote{R source code: \url{https://github.com/yenchic/Morse_Smale}}.
Note that one can use
the R package `{\sf msr}\footnote{\url{https://cran.r-project.org/web/packages/msr/index.html}}' that 
to perform
data analysis with the Morse-Smale complex (see \cite{gerber2010visual,gerber2011data,gerber2013morse} for more details).

\begin{figure}
\center
\includegraphics[width=2in]{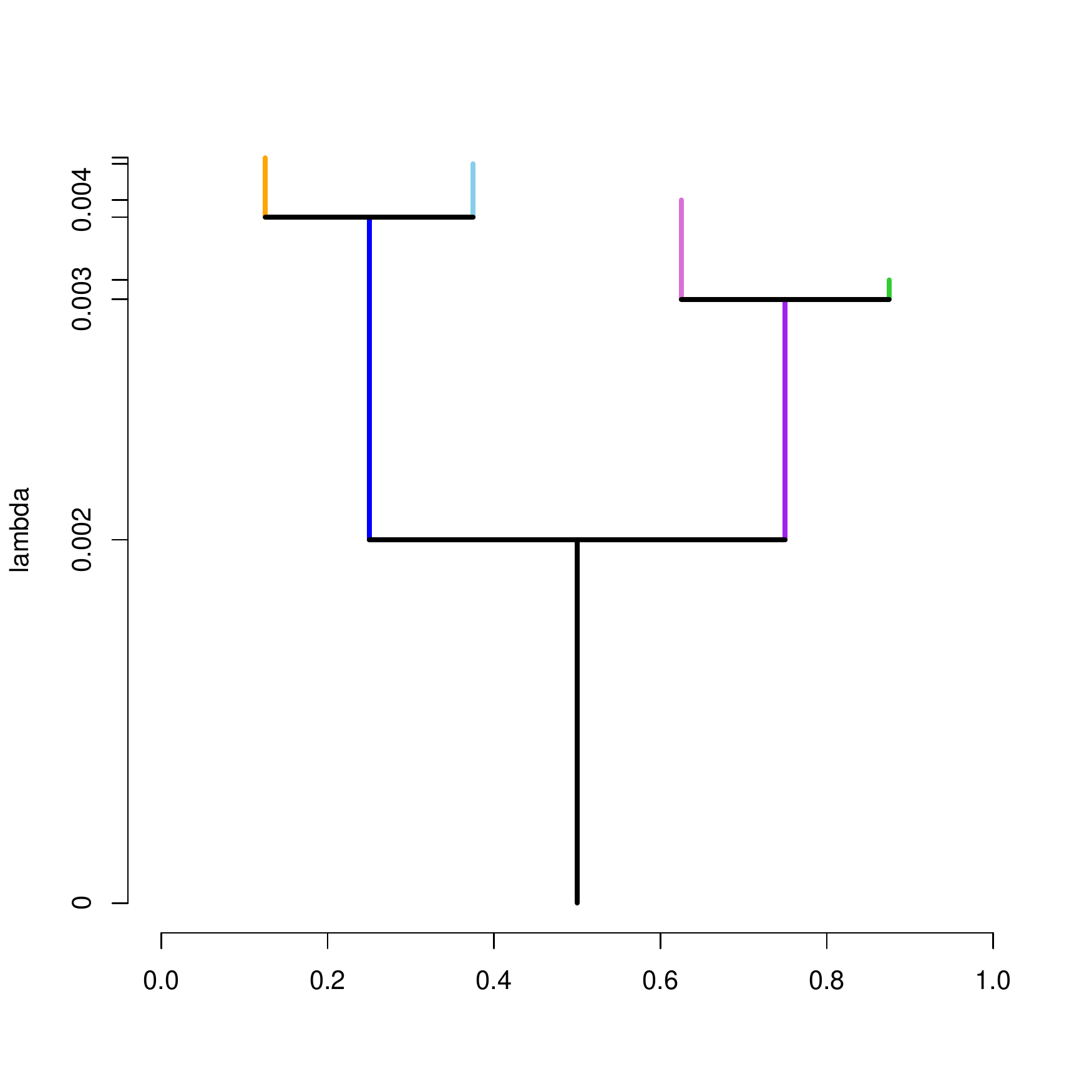}
\includegraphics[width=2in]{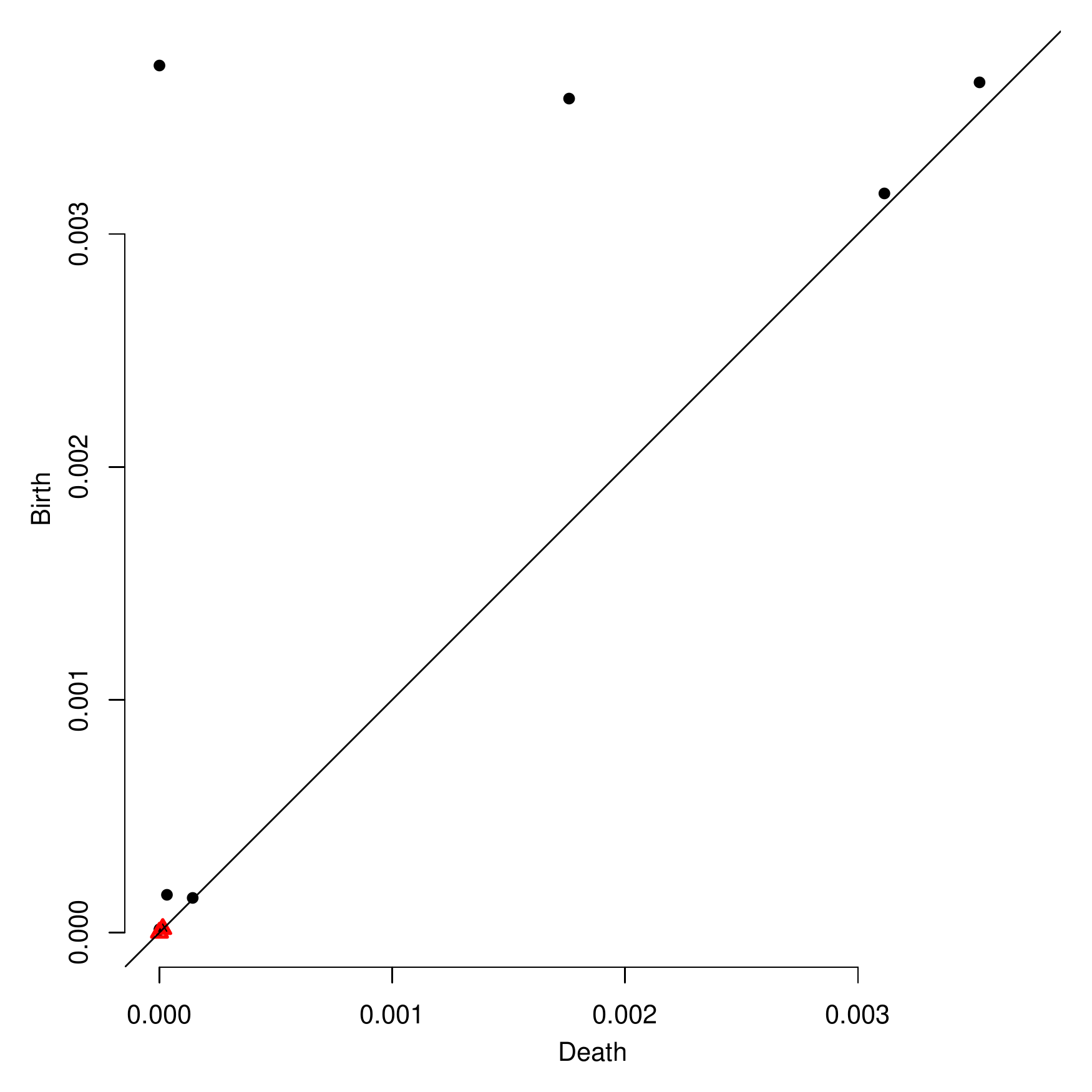}
\caption{Estimating topological features of KDE. 
This is the same dataset as in the right panel of Figure~\ref{fig::ex0}.
{\bf Left:} 
Cluster tree, a tree structure representing KDE.
The four leaves correspond to the the four high-density local modes in Figure~\ref{fig::geo1}
(other local modes are tiny so the cluster tree algorithm ignores them).
{\bf Right:} 
The persistent diagram. 
The top four black dots denote the four persistent topological features (four connected components)
of KDE, which are created by the four high-density local modes in Figure~\ref{fig::geo1}. 
There are several structures in the bottom left corner; they correspond to
the topological noises in constructing a persistent diagram.
For more details, we refer the reader to \cite{wasserman2016topological}.
}
\label{fig::geo2}
\end{figure}

\subsection{Cluster Trees}

Cluster trees (also known as density trees \cite{stuetzle2003estimating, klemela2009smoothing})
are tree-structured objects summarizing the structure of the underlying density function.
The left panel of Figure \ref{fig::geo2} provides an example of a cluster tree corresponding 
to KDE in the right panel of Figure~\ref{fig::ex} (also the same dataset as in Figure~\ref{fig::geo1} and \ref{fig::geo2}). 
A cluster tree is constructed as follows.
Recall that $L_\lambda =\{x: p(x)\geq \lambda\}$ is the density level set at the level $\lambda$.
When the level $\lambda$ is too large,  $L_\lambda$ will be empty, because
no region has a density value above such level.
When we gradually decrease $\lambda$ from a large number,
at some levels (when $\lambda$ hits the density value of local modes), 
a new connected component will be created; this corresponds the creation of a connected component. 
Moreover, at particular levels, two or more connected components will merge (generally
at the density value of local minima or saddle points); this corresponds to the elimination of a connected component.
Cluster tree uses a tree structure to summarize the creation and elimination of
connected components at different levels. 
Since a cluster tree always live in a 2D plane, it is an excellent tool
for visualizing a multivariate density function. 
For more details, we refer to the review paper in \cite{wasserman2016topological}.

The above defines a cluster tree using the underlying population density $p$.
In practice, we can construct an estimated cluster tree using KDE $\hat{p}_n$.
Convergence of the cluster tree estimator was studied in 
\cite{balakrishnan2013cluster,chaudhuri2014consistent,eldridge2015beyond,chen2016generalized}.
\cite{jisu2016statistical} provides a procedure for constructing
confidence sets of cluster trees based on bootstrapping KDE $\hat{p}_n$. 
In R, one can use 
the package `{\sf TDA}\footnote{\url{https://cran.r-project.org/web/packages/TDA/index.html}}'
to construct a tree estimator of KDE.

\subsection{Persistent Diagram}

A persistent diagram \citep{cohen2007stability,edelsbrunner2012persistent,wasserman2016topological} 
is a diagram summarizing the topological features of a density function $p$.
The construction of a persistent diagram is very similar to that of a cluster tree, but now
we focus on not only the connected components but also higher-order topological structures,
such as loops and voids (see \cite{fasy2014confidence, wasserman2016topological}
for a more details). 

There are several means of estimating a persistent diagram.
A natural approach is to use the persistent diagram of KDE $\hat{p}_n$.
For such an estimator, the stability theorem in \cite{cohen2007stability}
together with the uniform convergence in Section~\ref{sec::CR}
are sufficient to
prove the convergence of the estimated persistent diagram
toward the population persistent diagram. 
For statistical inference, \cite{fasy2014confidence} 
proposes a bootstrap approach over KDE
to construct a confidence set.
In practice, one can use R package `{\sf TDA}\footnote{\url{https://cran.r-project.org/web/packages/TDA/index.html}}' 
to construct the persistent diagram of KDE.

The right panel of Figure \ref{fig::geo2} shows an example of the persistent diagram
of connected components (zeroth-order topological features) and loops (first-order topological features)
of KDE described in the right panel of Figure~\ref{fig::ex}. 
At the top of the figure, the four dots indicate the existence of the four high-density local modes.
In the bottom left regions, the black dots and red triangles are
the topological noises representing low-density local modes (black dots)
and low-density ``loops" structures (red triangles; first-order topological features). 
Note that the two low-density local modes (black crosses) in the right part of both panels of Figure~\ref{fig::geo1}
are topological noises corresponding to the two black dots in the bottom left corner of the persistent diagram.

%
%



\section{Estimating the CDF}	\label{sec::CDF}

KDE can also be used to estimate a CDF
\citep{nadaraya1964some}.
The estimator is simple; we just integrate KDE (for simplicity, here we consider the univariate case):
\begin{equation}
\hat{F}_n(x) = \int^x_{-\infty} \hat{p}_n(y)dy.
\label{eq::CDF}
\end{equation}
Convergence of such estimators was extensively analyzed 
soon after their introduction \citep{winter1973strong, azzalini1981note, reiss1981nonparametric, fernholz1991almost,yukich1992weak, mack1984remarks}. 

Similarly to the pointwise error rate, one can show that 
$$
\hat{F}_n(x)-F(x) = O(h^2)+O_P\left(\sqrt{\frac{1}{n}}\right)+O_P\left(\sqrt{\frac{h}{n}}\right)
$$
(see the derivation in \cite{nadaraya1964some} and \cite{azzalini1981note}).
Again, the first quantity $O(h^2)$ is related to the bias.
The other two quantities are related to the stochastic variation. 
Under such a rate, the optimal smoothing bandwidth will be $h^* \asymp n^{-1/3}$, which leads to the error rate
$$
\hat{F}_n(x)-F(x) = O_P\left(\sqrt{\frac{1}{n}}\right),
$$
the same as using the empirical CDF. 
Note that as long as $h = O(n^{-1/4})$, we will obtain the square root error rate.

To construct a confidence band of $F(x)$ via $\hat{F}_n(x)$,
one can use the 
uniform central limit theorem proposed by
\cite{gine2008uniform} to relate the uniform loss to the supremum of a Gaussian process
and then use either the limiting distribution or the bootstrap.
Note that to apply the result in \cite{gine2008uniform},
one have to undersmooth (see Theorem 4 and Section 4.1.1 in \cite{gine2008uniform}) the data
or use a higher order kernel function (see Remark 7 and Corollary 2 in \cite{gine2008uniform}). 

\subsection{ROC Curve}

KDE can also be applied to estimate and infer the receiver operating characteristic (ROC) curve \citep{mcneil1984statistical}.
In the setting of the ROC curve, we observed two samples.
The first is the sample of healthy subjects, whose responses are $X_1,\cdots,X_n$ from an unknown density $P$. 
The other is the sample of diseased individuals, whose response are $Y_1,\cdots,Y_m$ from an unknown density $G$. 
Consider a simple rule of classification based on
choosing a cutoff point $s$ such that 
we classify an individual as diseased if its response value is larger than $s$, otherwise it is classified as a healthy individual. 

For such a rule, the sensitivity is $SE(s) = 1-G(s)$, the probability 
of detecting a diseased subject. 
We also define the specificity $SP(s) = F(s)$, the probability 
of successfully assigning a healthy subject to the healthy group. 
Then, the ROC curve is defined as the plotting of
the true positive fraction $SE(s)$ versus the false positive fraction $1-SP(s)$, or
equivalently, as plotting the function
$$
ROC(t) = 1 -G(F^{-1}(1-t)).
$$
A recent review on ROC curves can be found in \cite{demidenko2012confidence}.

A classical nonparametric approach of estimating $ROC(t)$
is to plug-in the empirical CDF estimator for both $F$ and $G$ \citep{hsieh1996nonparametric}.
As an alternative, 
one can use the integrated KDEs of both samples to estimate the ROC curve
\citep{zou1997smooth,zhou2002comparison, hall2004nonparametric}
as equation \eqref{eq::CDF}.
This is often called a smoothed estimator of the ROC curve because
the resulting ROC curve estimator is generally a smooth curve. 

To construct confidence bands of an ROC curve,
most methods propose using the plug-in estimate 
from the empirical distribution
and constructing the confidence band by the bootstrap
\citep{moise1985comparison,campbell1994advances,macskassy2004confidence}.
A formal proof of the theoretical validity of such a bootstrap approach is provided in
\cite{hall2004nonparametric,horvath2008confidence,bertail2009bootstrapping}.
Note that can also construct a confidence band by bootstrapping the 
smoothed ROC curve estimator and using the method proposed in 
Section~\ref{sec::UCS} to construct a confidence band.

\section{Conclusion and Open Problems}	\label{sec::future}

In this tutorial, we reviewed KDE's basic properties and 
its applications in estimating structures of the underlying density function. 
For readers who would like to learn more about different varieties of KDE,
we recommend
\cite{wasserman2006all} and \cite{scott2015multivariate}.
Because this is a tutorial, we ignore many advanced
topics such as the minimax theory and adaptation.
An introduction of these theoretical properties can be found in
\cite{tsybakov2009introduction}.

Although KDE has been widely studied since its introduction in the 1960s, 
there are still open problems that deserve further investigation. 
Here we briefly discuss some open problems related to the materials in this tutorial.


\begin{itemize}
\item {\bf Confidence bands of other KDE-type estimators.}
In addition to estimating a probability density function and its related structures, 
the idea of kernel smoothing can be applied to estimate a regression,
hazard, or survival function.
Moreover, in casual inference, we might be interested in the difference between the regression/hazard/survival function 
from the control group and that from the treatment group as a characteristic of the treatment effect.
One example is the conditional average treatment effect 
\citep{lee2009nonparametric, hsu2013consistent, ma2014treatment,abrevaya2015estimating}.
Although in this tutorial we have seen methods of constructing confidence bands of density functions, 
how to construct a (asymptotically) valid confidence band of these 
functions remains an open question.

\item {\bf Multidimensional problems.}
When the dimension of the data $d$ is large,
KDE poses several challenges. 
First, KDE (and most other nonparametric density estimators) suffers severely
from the so-called \emph{curse of dimensionality}:
The optimal convergence rate $O(n^{-\frac{2}{d+4}})$ is very slow when $d$ is large,
and this slow convergence rate cannot be improved \citep{stone1982optimal,tsybakov2009introduction} 
unless we assume extra smoothness.
One way to solve this problem is to find density surrogates that can be estimated easily
and to switch our parameter of interest to a density surrogate.
However, this rises the question of what the correct surrogates and the corresponding estimators
are, and this
still remains unclear.
Another issue of KDE in multi-dimensions is visualization.
When $d>3$, we can no longer see the entire KDE, and therefore we must use visualization tools
to explore our density estimates. 
However, it is still unclear how to choose a visualization tool in practice.

\item {\bf More about geometric/topological structures.}
In Section~\ref{sec::GT}, we saw that several useful geometric and topological structures 
can be estimated by the corresponding structures of KDE. 
However, we do not yet fully understand the behavior of these estimators.
For instance, how to choose the smoothing bandwidth that optimally estimate these structures
is unclear.
Handling this issue may require generalizing the concept of the MISE to the set estimator \citep{chen2015optimal}
and choosing the smoothing bandwidth that minimizes such an error measurement.
In addition to bandwidth selection, 
uniform inference remains an open question for these structures. 
Although there are methods of constructing confidence sets of most of these structures,
it is unclear whether the resulting confidence sets are uniform for a collection of density functions. 
Moreover, 
theoretical optimality, such as the minimax theory, remains unclear for several of these structures,
presenting another set of open questions in the study of KDE.

\end{itemize}


\section*{Acknowledgement}
We thank two referees for the very useful suggestions.
We also thank Gang Chen, Aurelio Uncini, and Larry Wasserman for useful comments.

\bibliographystyle{abbrvnat}
\bibliography{tutorial_kde.bib}



\section*{Supplementary material}

Here is a simple illustration on how to compute the kernel density estimator (KDE) using R. We use the dataset `geyser` in the package `MASS` and focus on the variable `waiting`. We will not discuss the details of each function but only the important arguments.

### 1D case

```r
library(MASS)
summary(geyser)
```

```
##     waiting          duration
##  Min.   : 43.00   Min.   :0.8333
##  1st Qu.: 59.00   1st Qu.:2.0000
##  Median : 76.00   Median :4.0000
##  Mean   : 72.31   Mean   :3.4608
##  3rd Qu.: 83.00   3rd Qu.:4.3833
##  Max.   :108.00   Max.   :5.4500
```

```r
data1d = geyser[,1]
```

To use the KDE of a univariate dataset, we use the built-in function `density`:

```r
h1=4
data1d_kde = density(data1d, bw=h1, from=40, to=110)
  ## bw = h1: we set the smoothing banwdith = h1.
  ## from = 40, to = 110: the range we are evaluating the density is from 40 to 110.

##### make a plot
par(mar=c(4,4,2,1))
plot(data1d_kde, lwd=4, col="blue",ylim=c(0, 0.045), cex.axis=1.5,
     main="Kernel Density Estimator", cex.lab=2, cex.main=1.5, ylab="")
mtext("Density", side=2, line=2.2, cex=2)
```

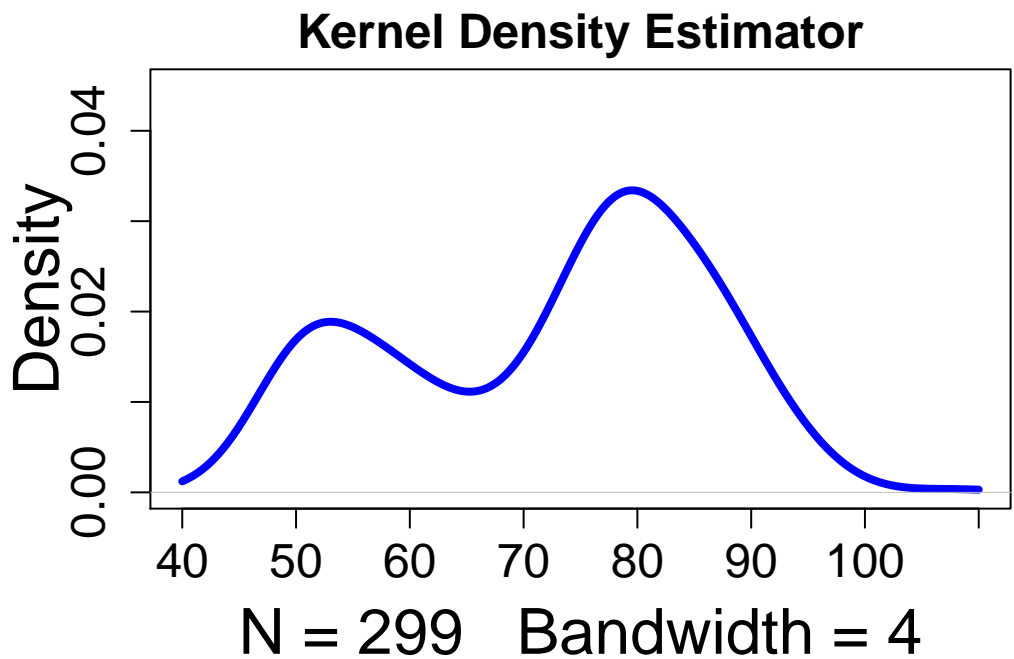

You can use the library `ks` to choose the smoothing bandwidth. The function `hns()` (and `Hns()` for

multivariate case) choose the smoothing bandwidth according to the Silverman's rule. The function `hlscv()` (and `Hlscv()`) uses the least squre cross-validation approach. The function `hpi()` (and `Hpi()`) applies the plug-in method for bandwidth selection.

```r
library(ks)
h2 = hns(data1d)
h3 = hlscv(data1d)
h4 = hpi(data1d)
data1d_kde2 = density(data1d, bw=h2, from=40, to=110)
data1d_kde3 = density(data1d, bw=h3, from=40, to=110)
data1d_kde4 = density(data1d, bw=h4, from=40, to=110)
h2
```

```
## [1] 4.705068
```

```r
h3
```

```
## [1] 2.24592
```

```r
h4
```

```
## [1] 2.787587
```

```r
##### make a plot
par(mar=c(4,4,2,1))
plot(data1d_kde2, lwd=4, col="dodgerblue",ylim=c(0, 0.045), cex.axis=1.5,
     main="Kernel Density Estimator", cex.lab=2, cex.main=1.5, ylab="",
     xlab="")
lines(data1d_kde3, lwd=4, col="limegreen")
lines(data1d_kde4, lwd=4, col="brown")
mtext("Density", side=2, line=2.2, cex=2)
legend("topleft",c("hns()","hlscv()","hpi()"), col=c("dodgerblue","limegreen","brown"), lwd=5)
```

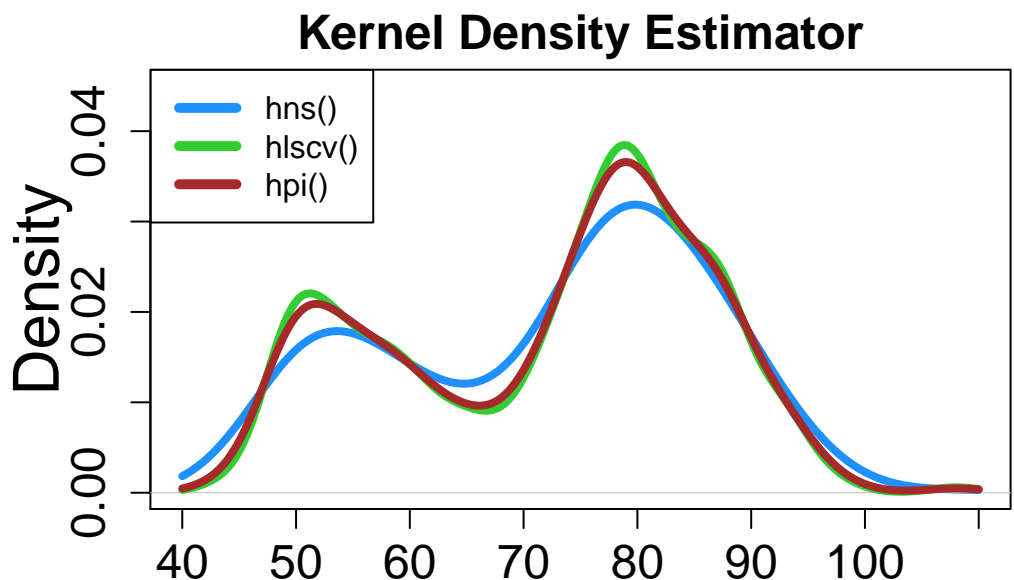

**Confidence intervals**

Here we present how one may construct a confidence interval using method 1 (plug-in approach) and method 3 (bootstrap approach). Let the significance level $\alpha = 0.05$. Note that for the Gaussian kernel, $\mu_K = \frac{1}{2\sqrt{\pi}}$.

2

The following is how we use the plug-in approach to construct a confidence interval:

```
alpha0=0.05
n1 = length(data1d)
t0 = qnorm(1-alpha0)*sqrt(1/(2*sqrt(pi))/(n1*h1)*data1d_kde$y)
  ## data1d_kde$y: the density value evaluated at each point of data1d_kde$x.

##### make a plot
par(mar=c(4,4,2,1))
plot(data1d_kde, lwd=3, col="tan4", ylim=c(0, 0.045), cex.axis=1.5,
     main="Confidence Interval (PI)", xlab="", ylab="", cex.main=1.5)
mtext("Density", side=2, line=2.2, cex=2)
polygon(c(data1d_kde$x, rev(data1d_kde$x)),
        c(data1d_kde$y+t0, rev(data1d_kde$y-t0)),
        border="tan", col="tan")
abline(h=0, lwd=2)
lines(data1d_kde$x, data1d_kde$y, lwd=3, col="tan4")
```

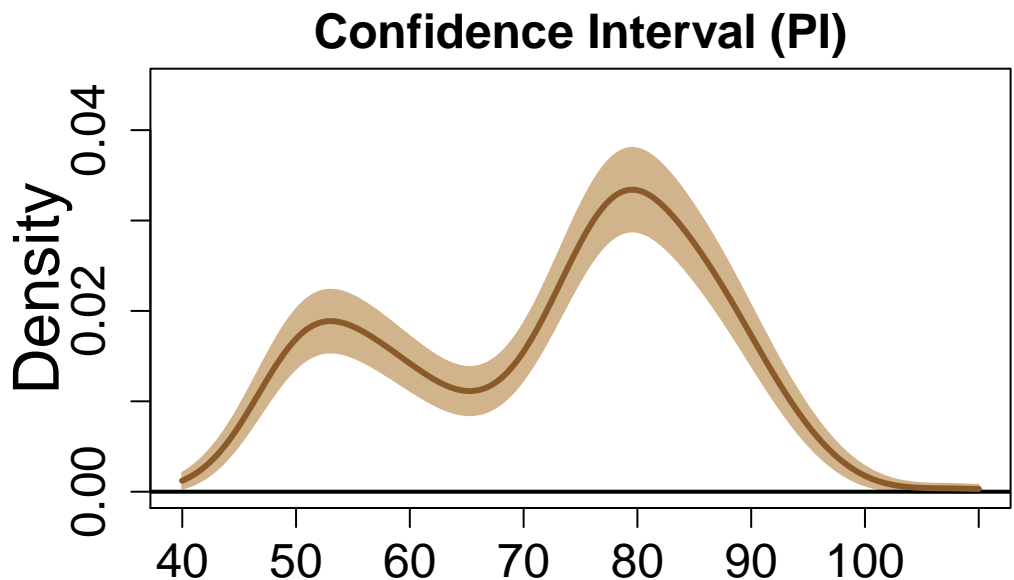

For the case of bootstrap approach, here is the procedure:

```
n_BT = 1000
  ## number of bootstrap samples
kde_seq_BT_m = matrix(NA, nrow=n_BT, ncol= length(data1d_kde$y))
for(j in 1:n_BT){
  data1d_BT = data1d[sample(n1, n1, replace=T)]
  data1d_kde_BT = density(data1d_BT, bw=h1, from=40, to=110)

  kde_seq_BT_m[j,] = abs(data1d_kde_BT$y-data1d_kde$y)
}
  ## each row of 'kde_seq_BT_m' contains one bootstrap difference

t_pt = rep(0, length(data1d_kde$y))
for(l in 1:length(data1d_kde$y)){
  t_pt[l] = quantile(kde_seq_BT_m[,l], 1-alpha0)
}
  ## t_pt: the 1-\alpha quantile of deviation at each point
```

```
##### make a plot
par(mar=c(4,4,2,1))
plot(data1d_kde, lwd=3, col="blue", ylim=c(0, 0.045), cex.axis=1.5,
     main="Confidence Interval (Bootstrap)", xlab="", ylab="", cex.main=1.5)
mtext("Density", side=2, line=2.2, cex=2)
polygon(c(data1d_kde$x, rev(data1d_kde$x)),
        c(data1d_kde$y+t_pt, rev(data1d_kde$y-t_pt)),
        border="skyblue", col="skyblue")
abline(h=0, lwd=2)
lines(data1d_kde$x, data1d_kde$y, lwd=3, col="blue")
```

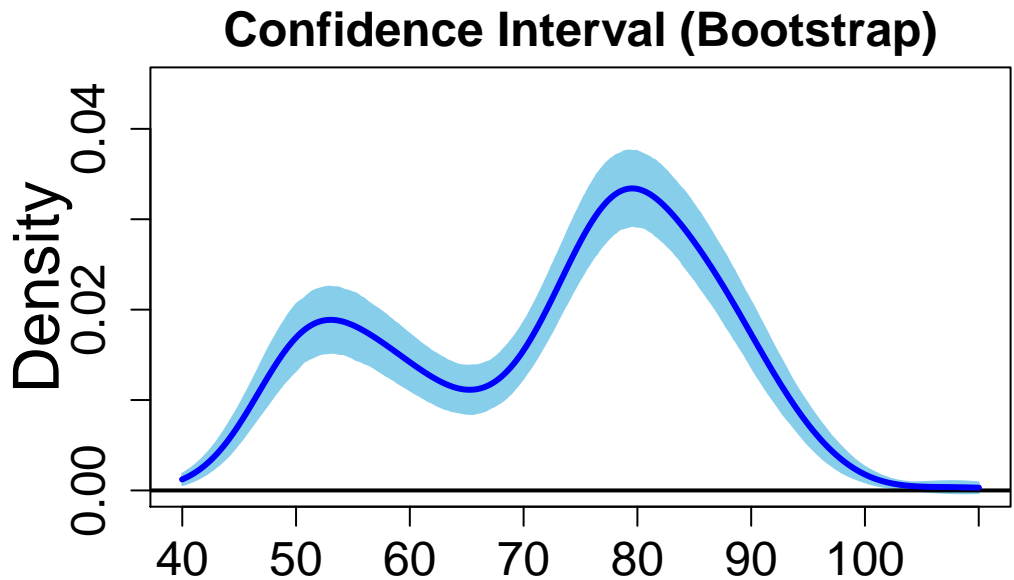

**Confidence bands**

Now we present the implementation of confidenc bands. We will focus on two approaches: the traditional bootstrap approach and the bootstrapping with the debiased KDE approach, which is presented in Section 3.3.3.

The implementation of the bootstrap confidence bands is very similar to the bootstrap confidence intervals. The main difference is that we replace the pointwise difference by the uniform difference. In more details:

```
kde_seq_sup = rep(0, n_BT)
for(j in 1:n_BT){
  data1d_BT = data1d[sample(n1, n1, replace=T)]
  data1d_kde_BT = density(data1d_BT, bw=h1, from=40, to=110)
    ## bootstrap KDE
  kde_seq_sup[j] = max(abs(data1d_kde_BT$y-data1d_kde$y))
}
t_sup = quantile(kde_seq_sup, 1-alpha0)

##### make a plot
par(mar=c(4,4,2,1))
plot(data1d_kde$x, data1d_kde$y, lwd=3, col="purple", ylim=c(0, 0.045), cex.axis=1.5,
     main="Confidence Band", xlab="", ylab="", type="l", cex.main=1.5)
mtext("Density", side=2, line=2.2, cex=2)
polygon(c(data1d_kde$x, rev(data1d_kde$x)),
```

4

```r
        c(data1d_kde$y+t_sup, rev(data1d_kde$y-t_sup)),
        border="plum1", col="plum1")
abline(h=0, lwd=2)
lines(data1d_kde$x, data1d_kde$y, lwd=3, col="purple")
```

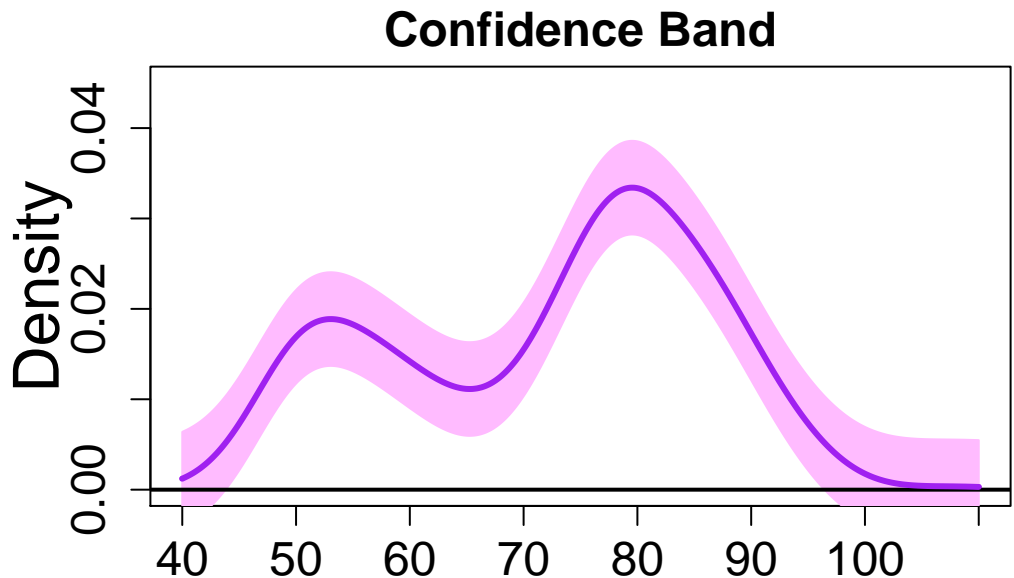

Now we present the implementation of bootstrapping the debiased KDE approach. Recall that the debiased KDE is:

$$\tilde{p}_n(x) = \hat{p}_n(x) - \frac{h^2}{2} \cdot c_K \cdot \hat{p}_n''(x)$$

in the univariate case. The constant $c_K = 1$ for the Gaussian kernel. Thus, we construct the confidence band by the following:

```r
library(ks)
  ## this library provides density derivative estimation
X0_kdde = kdde(data1d, h=h1, eval.points = data1d_kde$x, deriv.order = 2)
  ## eval.points = data1d_kde$x: we evalute the result at data1d_kde$x
  ## deriv.order = 2: we are computing the second derivative
de_kde_seq = data1d_kde$y-0.5*h1^2*X0_kdde$estimate
  ## this is the debiased KDE, evaluated at points of data1d_kde$x

de_kde_seq_sup = rep(0, n_BT)
for(j in 1:n_BT){
  data1d_BT = data1d[sample(n1, n1, replace=T)]
  data1d_kde_BT = density(data1d_BT, bw=h1, from=40, to=110)
  data1d_kdde_BT = kdde(data1d_BT, h=h1, eval.points = data1d_kde$x, deriv.order = 2)
  de_kde_seq_BT = data1d_kde_BT$y-0.5*h1^2*data1d_kdde_BT$estimate
    ##  the bootstrap debiased KDE
  de_kde_seq_sup[j] = max(abs(de_kde_seq-de_kde_seq_BT))
}
t_de_sup = quantile(de_kde_seq_sup, 1-alpha0)

##### make a plot
par(mar=c(4,4,2,1))
plot(data1d_kde$x, de_kde_seq, lwd=3, col="seagreen", ylim=c(0, 0.045), cex.axis=1.5,
     main="Confidence Band (Debiased)", xlab="", ylab="", type="l", cex.main=1.5)
mtext("Density", side=2, line=2.2, cex=2)
```

```
polygon(c(data1d_kde$x, rev(data1d_kde$x)),
        c(de_kde_seq+t_de_sup, rev(de_kde_seq-t_de_sup)),
        border="palegreen", col="palegreen")

abline(h=0, lwd=2)
lines(data1d_kde$x, de_kde_seq, lwd=3, col="seagreen")
```

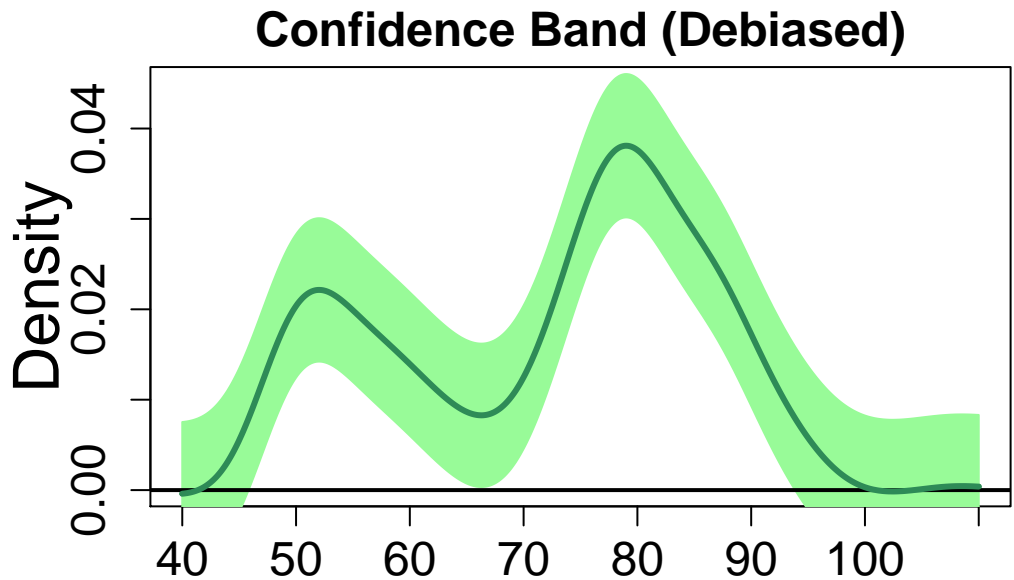

### 2D case

In the bivariate case, we recommend to use the function `bkde2D` in the package `KernSmooth`. The two variables being used are the current waiting time and the next waiting time.

```
library(KernSmooth)
```

```
## KernSmooth 2.23 loaded
## Copyright M. P. Wand 1997-2009
```

```
data2d = cbind(data1d[1:(n1-1)], data1d[2:n1])
  ## getting the current waiting time and the next waiting time

##### make a plot
par(mar=c(4,4,2,1))
plot(data2d, pch=16, xlab="Current waiting time", ylab="",
     cex.axis=1.5, cex.lab=2)
mtext("Next waiting time", side=2, line=2.2, cex=2)
```

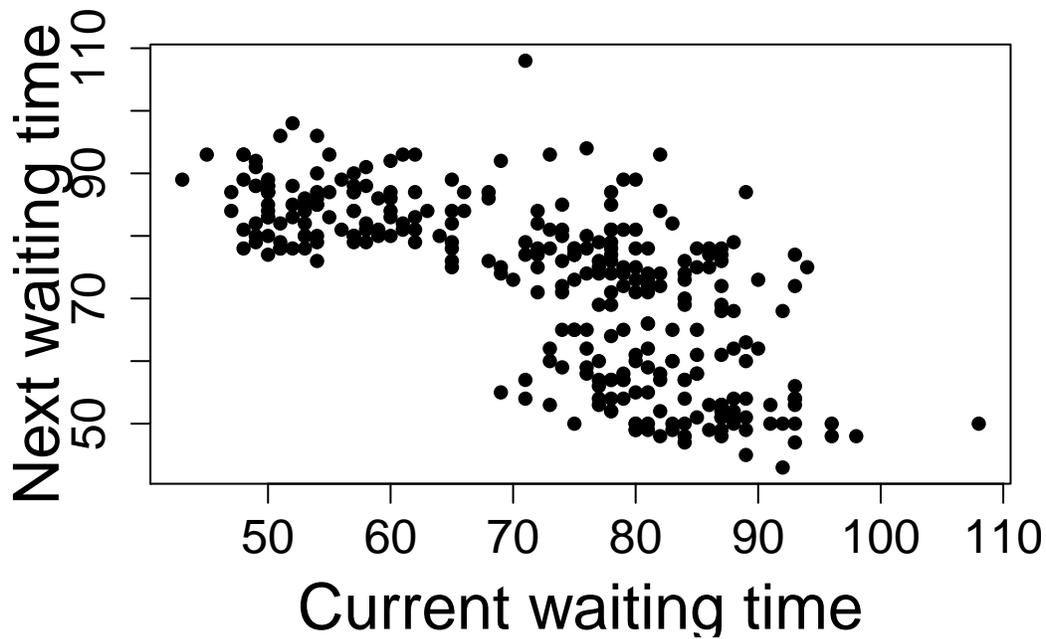

To illustrate the 2D KDE, we use the function `contour` which displays the contour plot. We choose the smoothing bandwidth to be 5.5 (this is just an arbitary choice).

```
h2 = 5.5
data2d_kde = bkde2D(data2d, bandwidth = h2)

##### make a plot
par(mar=c(4,4,2,1))
contour(data2d_kde$x1,data2d_kde$x2,data2d_kde$fhat, lwd=2, col="blue",
        xlab="Current waiting time", ylab="", cex.axis=1.5,
        cex.lab=2)
mtext("Next waiting time", side=2, line=2.2, cex=2)
points(data2d, pch=16)
```

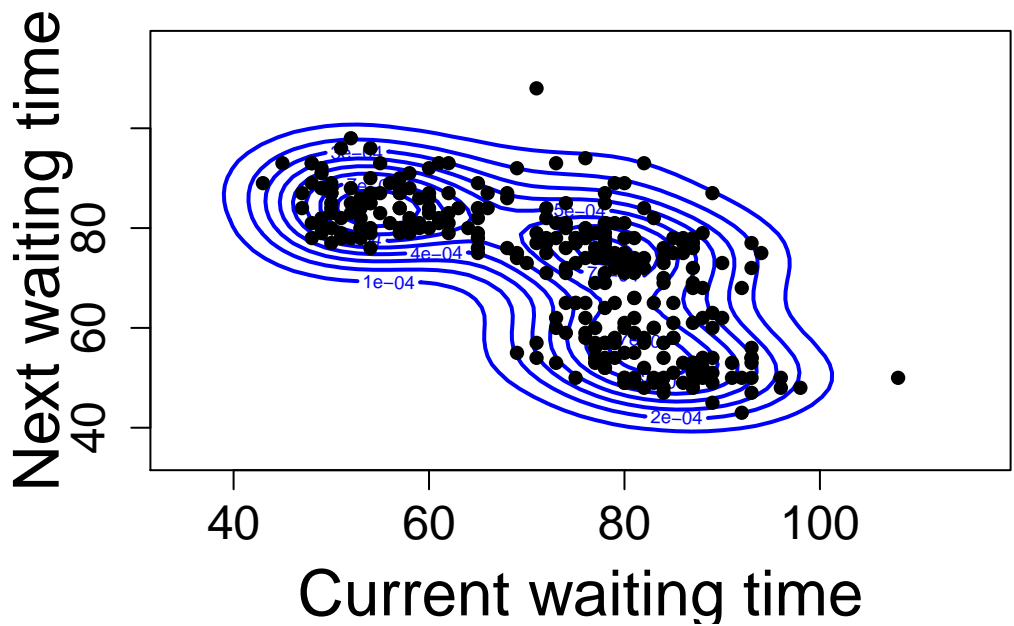

**Mode clustering**

To find local modes and perform mode clustering, we use the mean shift algorithm in the package `LPCM`:

```
library(LPCM)

##### make a plot
par(mar=c(4,4,2,1))
data2d_ms = ms(data2d, h=h2, scaled=F, cex.axis=1.5,
               xlab="Current waiting time", ylab="", cex.lab=2)
  ## scaled = F: this will not scale the data by its range
mtext("Next waiting time", side=2, line=2.2, cex=2)
```

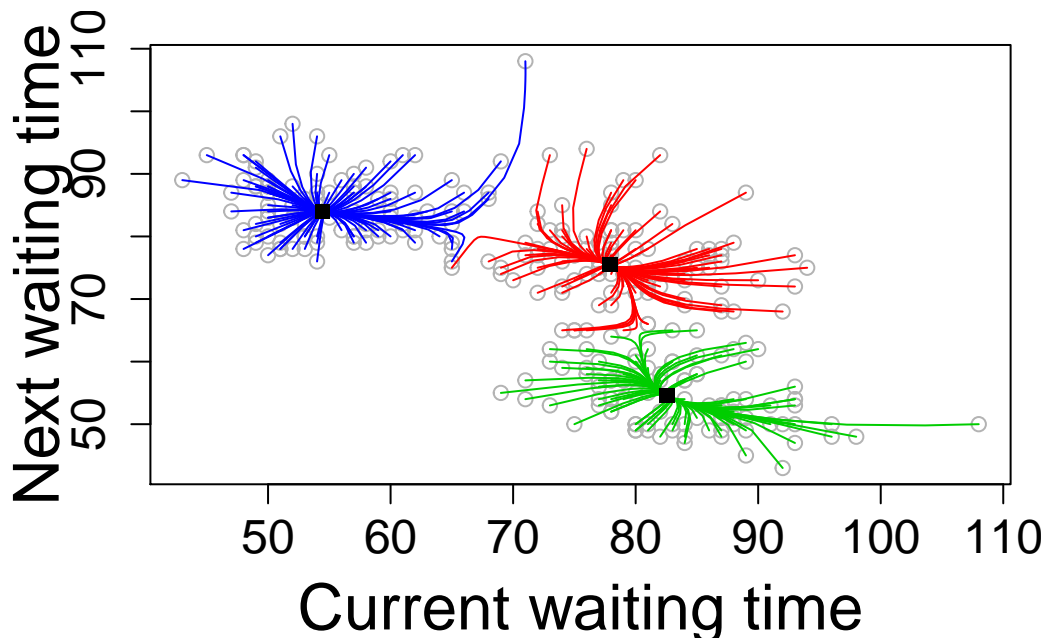

Note that the function `ms` will automatically plot the result; to disable this function, set the argument `plotms=0`. The colored curves are gradient flow lines starting at each data point. Different colors denote gradietnt flow lines with different destinations. The 3 destinations (local modes) are marked by black boxes. Based on the destination of the gradient flows, we partition data points into 3 clusters, denoted by the color of each flow line.

**Cluster tree and persistent diagram**

Our final demontration is the cluster tree and the persistent diagram. Both can be computed using the package `TDA`. For the cluster tree, we use the function `clusterTree`:

```
library(TDA)
data2d_cl = clusterTree(data2d, density="kde", h=h2, k=10)
  ## k: to compute the connected component, we need to assign this neighborhood constant.
  ##    generally it is chosen to be 10-20.

##### make a plot
par(mar=c(4,4,2,1))
plot(data2d_cl, col=c("black","blue","purple","red","limegreen"))
```

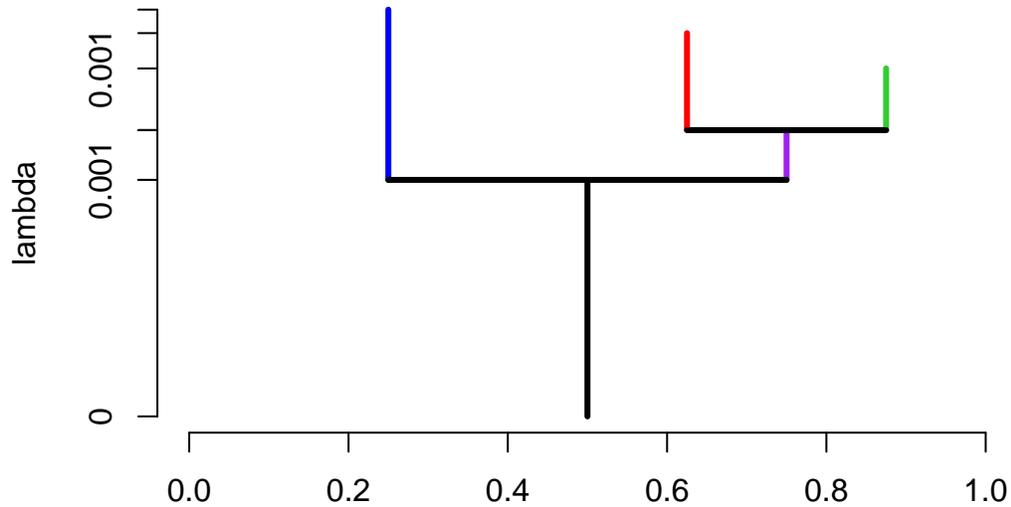

The X-axis is just an axis showing the tree structures so its numeric value does not have any meaning. The Y-axis is about the level sets at certain density levels. The top three branches are the three local modes and the corresponding connected components around them. When we move down the density threshold $\lambda$, every time $\lambda$ pass the density of a local mode, a new connected component will be created. And when we keep moving down $\lambda$, connected components will grow, and eventually merge with others when $\lambda$ pass certain levels (generally this will be density value of local minima or saddle points).

The purple branch appears when the connected components represented by red and green branches merged and it disappears when merging with the connected component represented by the blue branch.

Next, we demonstrate how to compute the persistent diagrma of the KDE using `TDA` package. We will use the function `gridDiag`, which first generates a grid and then evaluate the density on the grid and compute topological features.

```
xlim0 = c(20, 150)
ylim0 = c(20, 150)
  ## set the lim of the grid
data2d_pd = gridDiag(data2d, kde, h=h2, lim=cbind(xlim0, ylim0), by=2, sublevel=F)
  ## lim = ... : the limit of the grid.
  ## by = 2: the separation between grid points.
  ## sublevel = F: the density level sets are upper (super) level sets.

##### make a plot
par(mar=c(4,4,2,1))
plot(data2d_pd$diagram)
```

9

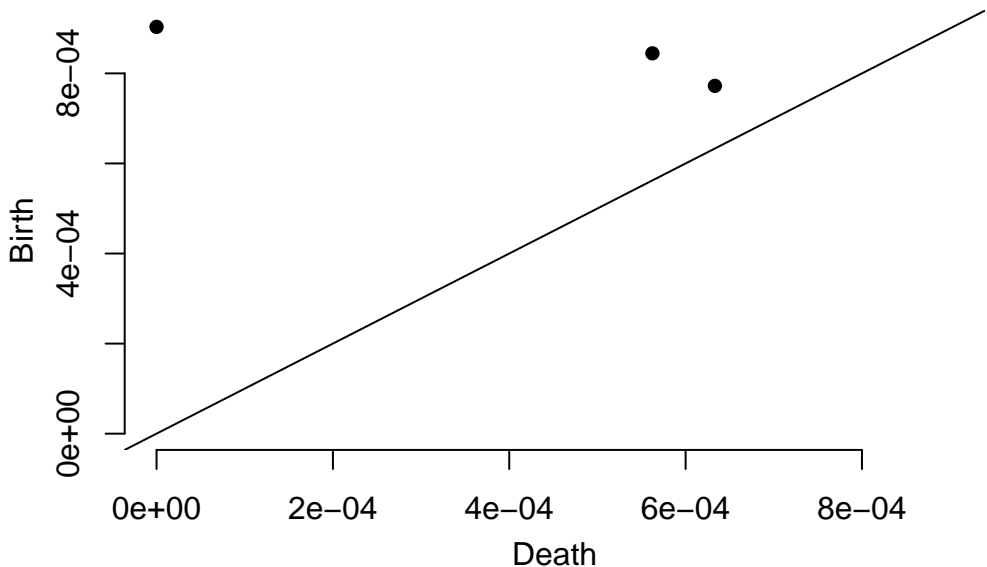

The three black dots denote the birth and death time of three connected components of the KDE. When we move down the level $\lambda$, the birth time is the level that a connected component is created and the death time is the level that a connected component is merged into others.

**Morse-Smale Complex**

Finally, we display a simple case where on can use the package `msr` to compute the Morse-Smale complex. Given a set of points (here we use the original data points but one can use a grid or any other collection of points), we first need to evaluate the density at these points. This first step can be done using the `kde()` function in the `TDA` pacakage. Having evaluated the density, we can use the function `msc.nn()` to obtain a partition of points.

```
data2d_density = TDA::kde(X=data2d,Grid=data2d,h=h2)
  ## Compute the KDE at each data point.
  ## We use "TDA::kde"" because 'ks' library also have the same function.
library(msr)
data2d_msc = msc.nn(y=data2d_density, x=data2d, knn=5)
  ## knn=5: the number of points to define a gradient flow on points

#### make a plot
par(mar=c(4,4,2,1))
col16 = rainbow(16)
col16 = col16[sample(16)]
  ## randomly permute the colors
contour(data2d_kde$x1,data2d_kde$x2,data2d_kde$fhat, lwd=1.5, col="gray30",
        xlab="Current waiting time", ylab="", cex.axis=1.5,
        cex.lab=2)
points(data2d, col=col16[data2d_msc$level[[1]]$partition], pch=16,
       cex=0.9)
  ## colors denote points belonging to different complices
mtext("Next waiting time", side=2, line=2.2, cex=2)
```

10

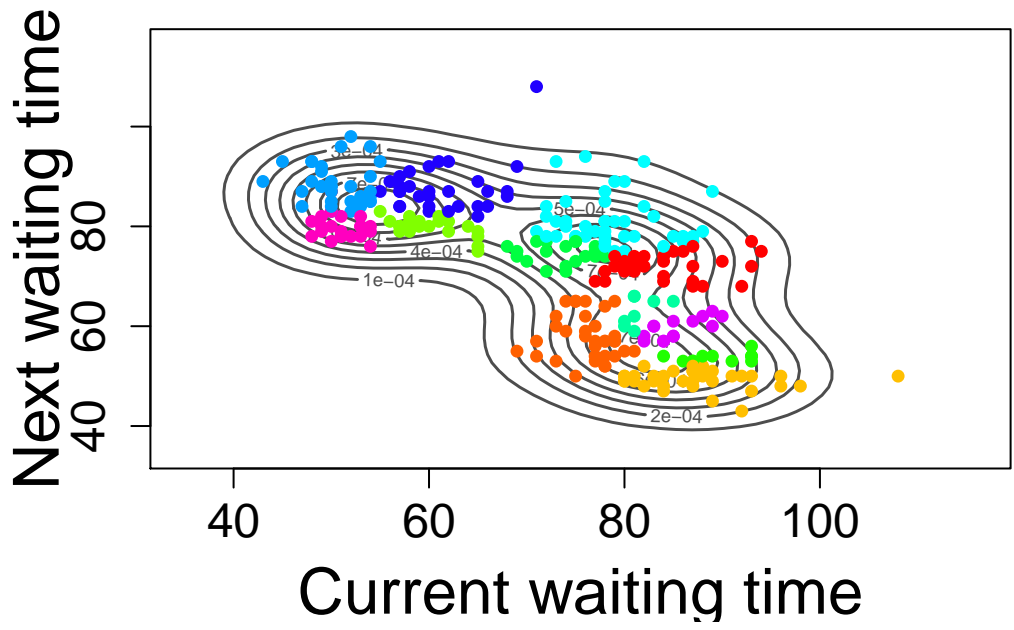

11



\end{document}